\documentclass[journal,twocolumn]{IEEEtran}
\usepackage{pifont}
\usepackage{cite}
\usepackage[dvips]{epsfig}
\usepackage{graphicx}
\usepackage{calligra}
\usepackage[T1]{fontenc}
\usepackage{bbold}
\usepackage{mathrsfs}

\usepackage{subfigure}
\usepackage{color}
\usepackage{amsmath}
\usepackage{cases}

\usepackage{amssymb}
\usepackage[justification=centering]{caption}
\usepackage{framed}

\usepackage{subfigure}
\usepackage{color}
\usepackage{amsmath}
\usepackage{bm}
\usepackage{amssymb}
\usepackage{amsbsy}
\usepackage{bbm}
\usepackage{dsfont}
\usepackage[normalem]{ulem}
\usepackage{xcolor,cancel}
\usepackage{enumerate}
\usepackage{paralist}

\usepackage{algorithm, algpseudocode}

\usepackage{amsthm}
\usepackage{setspace}

\newtheorem{theorem}{Theorem}

\newtheorem{condition}{Condition}

\newtheorem{corollary}{Corollary}

\newtheorem{definition}{Definition}

\newtheorem{proposition}{Proposition}

\newtheorem{lemma}{Lemma}

\newtheorem{remark}{Remark}

\newtheorem{claim}{Claim}

\newtheorem{assumption}{Assumption}

\newcommand{\cH}{\mathcal{H}}

\newcommand{\cQ}{\mathcal{Q}}
\newcommand{\cO}{\mathcal{O}}

\newcommand{\bx}{{\bf{x}}}

\newcommand{\ignore}[1]{{}}

\newcommand{\cbk}{\color{black}}

\newcounter{parentalgorithm}

\ifCLASSINFOpdf
\else
\fi

\begin{document}
%
\title{Distributed Detection over Blockchain-aided Internet of Things in the Presence of Attacks}

%
%
%

\author{Yiming Jiang  and Jiangfan Zhang, ~\IEEEmembership{Member,~IEEE}
\thanks{Y. Jiang and J. Zhang are with the Department of Electrical and Computer Engineering, Missouri University of Science and Technology, Rolla MO 65409 USA (e-mail: yjk7z@mst.edu, jiangfanzhang@mst.edu)} 
}


\maketitle

\begin{abstract}

Distributed detection over a blockchain-aided Internet of Things (BIoT) network in the presence of attacks is considered, where the integrated blockchain is employed to secure data exchanges over the BIoT as well as data storage at the agents of the BIoT. We consider a general adversary model where attackers jointly exploit the vulnerability of IoT devices and that of the blockchain employed in the BIoT. The optimal attacking strategy which minimizes the Kullback-Leibler divergence is pursued. It can be shown that this optimization problem is nonconvex, and hence it is generally intractable to find the globally optimal solution to such a problem. To overcome this issue, we first propose a relaxation method that can convert the original nonconvex optimization problem into a convex optimization problem, and then the analytic expression for the optimal solution to the relaxed convex optimization problem is derived. The optimal value of the relaxed convex optimization problem provides a detection performance guarantee for the BIoT in the presence of attacks. In addition, we develop a coordinate descent algorithm which is based on a capped water-filling method to solve the relaxed convex optimization problem, and moreover, we show that the convergence of the proposed coordinate descent algorithm can be guaranteed.

\end{abstract}


\IEEEpeerreviewmaketitle

\section{Introduction}
With the rapid development of smart devices and high-speed networks, the Internet of Things (IoT) has recently brought about an unprecedented increase in sensor resources and the deployment of sensor-like objects in safety-critical applications of vital societal interest, such as smart grids, healthcare informatics, manufacturing, and smart city \cite{al2015internet,lin2017survey}.

Typically, an IoT network consists of spatially distributed mutually-distrusting devices which sequentially generate and process exclusive data of a physical phenomenon of interest, and share their processed data with other devices over the network. In a conventional IoT (CIoT) network which is equipped with a cloud (or a fusion center), IoT devices transfer their data to a cloud where the IoT devices' data are stored and processed. Thereby the CIoT is vulnerable to a single point of failure since if the cloud does not function normally, the CIoT is paralyzed. The data stored in the cloud is also at risk of being modified or deleted by malicious attackers aiming to hack into the cloud. Moreover, the CIoT is vulnerable to some other security threats as well, including attacks on data exchanges between IoT devices and the cloud and impersonation of IoT devices. This has recently led to great interest in studying the vulnerability of the CIoT in various applications, see \cite{ mosenia2016comprehensive,pan2021iout,marano2008distributed,zhang2020asymptotically, vempaty2013distributed,   zhang2015Asymptotically, mothukuri2021federated,zhang2018attack } and the references therein. 

Blockchain technology, an emerging secure distributed database technology that revolutionizes the way information is secured, distributed, and shared, has attracted enormous attention in recent years due to the following vital components \cite{puthal2018blockchain, dinh2018untangling}: (i) a chronologically ordered sequence of blocks that are cryptographically linked to each other and are shared, stored and synchronized over a network; (ii) strong cryptography enabling secure data storage and secure data exchanges, and (iii) a mutual consensus protocol that enables verification and validation of the authenticity and the integrity of stored and exchanged data, and thus enables mutual trust over a network instead of relying on a central authority. 

By taking advantage of the security-by-design and distributed nature without needing any central authority, blockchain lately has been integrated into IoT networks, and this kind of newly emerging blockchain-aided IoT (BIoT) network has been applied to a great deal of security-related applications, such as smart grids \cite{zhuang2020blockchain,ferrag2019deepcoin}, vehicular networks \cite{liu2021blockchain,hassija2020traffic}, and smart city \cite{jiang2020blockchain,yang2021privacy}. Blockchain provides feasible solutions to address many common security threats to IoT networks. For example, in a blockchain network, by virtue of a consensus protocol, each node maintains a local copy of a blockchain which can be guaranteed to be identical to other nodes' copies. If one node's copy of the blockchain is maliciously corrupted, it can be retrieved from other nodes in the blockchain network, and hence a single point of failure can therefore be prevented. Moreover, the communications over a blockchain network are secured by the cryptographic algorithms of the blockchain network which can prevent attacks that manipulate either transmitted messages or the identities of senders, such as the man-in-the-middle attack and the Internet protocol address spoofing attack \cite{vempaty2013distributed, zhang2015Asymptotically, zhang2017functional, zhang2018approaches}. 

However, blockchain technologies cannot eradicate all security threats to the BIoT, and the vulnerability of the BIoT is determined by the vulnerability of IoT devices and that of the blockchain employed in the BIoT.  This is because on one hand, IoT devices in a BIoT are typically resource-limited and low-cost which makes them vulnerable to hacking. If hacked IoT devices in a BIoT deliberately transmit misleading data, then any task performed by the BIoT which  unwittingly utilizes the data from the hacked IoT devices can be significantly compromised. \cbk On the other hand, the data stored in a blockchain is not perfectly secure. As mentioned in \cite{nakamoto2008bitcoin,zaghloul2020bitcoin,karame2012two}, blockchains are prone to the double-spending attack (DSA), which is considered one of the most devastating attacks against blockchains. The DSA aims at falsifying the data which have already been stored in existing blocks of a blockchain. DSAs can be extremely injurious to blockchain networks. For example, one of the Bitcoin forks, Bitcoin Gold, suffered double-spending attacks in 2018, and again in 2020, with more than 17 million dollars lost in total. For a BIoT, a successful DSA can falsify the data stored in its blockchain without being perceived, and hence seriously compromise the data security of the BIoT. 

Detection problems which aim to distinguish between different hypotheses are prevalent among BIoT applications, such as intrusion detection \cite{liu2021blockchain,hu2019collaborative}, anomaly detection \cite{mothukuri2021federated,hassan2022anomaly}, and object detection\cite{jiang2020blockchain,jiang2021edge}. In a BIoT, the blockchain which stores the data from IoT devices are shared, distributed, and synchronized across different agents. Hence, the decision between different hypotheses can be distributedly made  at different agents which maintain local copies of the blockchain. Moreover, the decisions made by different agents can be guaranteed to reach a  consensus since different agents' local copies of the blockchain are the same owing to the consensus protocol of the blockchain.
In this paper,  we are interested in distributed detection over a BIoT in the presence of attacks which jointly exploit the vulnerability of IoT devices and that of the blockchain employed in the BIoT.

\subsection{Summary of Results and Main Contributions}
We consider a detection task over a BIoT which aims to make a decision between two hypotheses based on IoT devices' data stored in its blockchain. A general BIoT model is considered which is generalized from most existing works on BIoT applications where a blockchain is integrated into an IoT network to secure data storage and exchanges \cite{yang2021priscore,hjalmarsson2018blockchain,hassija2020traffic,lei2017blockchain, cha2018blockchain,lombardi2018blockchain,ling2019blockchain,li2018creditcoin,
zhou2018beekeeper,yang2021privacy}. It is worth mentioning that if the task of the BIoT application which makes use of data stored in the employed blockchain has to be accomplished by a time instant, such as the detection task considered in this paper, then from the perspective of this task, the blockchain employed in the BIoT should be considered finitely long. For example, for a practical detection task, a decision should be made within a limited period of time by using a limited amount of data stored in a finite number of blocks of the blockchain. Once the decision is made, the growth of the blockchain can be regarded as being terminated from the perspective of the detection task since the data stored in the blocks that are generated after the decision is made do not affect this detection task.

Based on the general BIoT model, we adopt the Kullback-Leibler divergence (KLD) as the performance metric, which is one of the most popular performance metrics for detection problems \cite{poor2013introduction}, and develop the detection performance guarantee for the BIoT in the presence of attacks which jointly exploit the vulnerability of IoT devices and that of the blockchain employed in the BIoT. To be specific, considering the attacks which jointly attack the IoT devices and the blockchain of a BIoT, we pursue the detection performance guarantee by minimizing the KLD between two hypotheses over all possible  malicious-data distributions. However, it can be shown that this minimization problem is nonconvex, and hence it is generally intractable to find the globally optimal solution for such a problem. To overcome this issue, we first propose a relaxation method to convert the nonconvex optimization problem into a convex optimization problem, and then the analytic expression for the optimal solution to the relaxed convex optimization problem is derived. In addition, we develop a coordinate descent algorithm which is based on a capped water-filling method to solve the relaxed convex optimization problem, and moreover, we show that the convergence of the proposed coordinate descent algorithm can be guaranteed. The optimal value of the relaxed convex optimization problem which is obtained from the proposed algorithm provides a performance guarantee for the BIoT in the presence of attacks.

\subsection{Related Work}\label{Related_work}
Recently, there has been great interest in integrating blockchain into diverse IoT applications to enhance data security, see \cite{zhuang2020blockchain,yang2021priscore,hjalmarsson2018blockchain,hassija2020traffic,lei2017blockchain, cha2018blockchain,lombardi2018blockchain,ling2019blockchain,li2018creditcoin,zhou2018beekeeper,yang2021privacy,shan2021poligraph} and the references therein. In \cite{li2018creditcoin}, the authors propose a blockchain-based vehicular announcement network where a blockchain is employed to protect data against tampering and make them widely available and accessible over the network. In \cite{hassija2020traffic}, a peer-to-peer vehicle network is proposed where traffic information is stored in a blockchain which secures information sharing. The blockchains in these works are considered as highly secure, immutable, and transparent distributed ledgers so that the data stored in the blockchains are assumed to be perfectly secure in these works. However, the data stored in a blockchain cannot be perfectly secured. For example, blockchains are vulnerable to the DSA which, if successful, can falsify the data stored in the blockchain without being perceived. In contrast, in this paper, we take into account the vulnerability of the blockchain employed in a BIoT when studying the performance of BIoT applications.

	The performance guarantee for distributed detection in the presence of attacks has been studied in various applications, see \cite{rawat2011collaborative,marano2008distributed,kailkhura2015distributeddetect,vempaty2013distributed} for instance. In \cite{marano2008distributed}, the KLD is adopted as the performance metric, and the authors develop the minimum KLD for distributed detection over a CIoT. The system and adversary models considered in \cite{marano2008distributed} are different from those considered in this paper, and it is shown that the KLD minimization problem in \cite{marano2008distributed} is a convex optimization problem. In contrast, we consider a distributed detection problem over a BIoT in the presence of attacks, and the minimization of the KLD for this problem can be shown to be nonconvex.  { In \cite{rawat2011collaborative}, the performance limit of collaborative spectrum sensing over a CIoT in the presence of attacks is analyzed where KLD is also adopted as the performance metric. A zero-sum game is employed in \cite{rawat2011collaborative} to model the interaction between a fusion center and an attacker.  Since decision makers are assumed to be unable to interact with attackers in the model considered in this paper, the game theory approach developed in \cite{rawat2011collaborative} cannot be applied to the problem considered in this paper.}


The paper is organized as follows. In section \ref{system_model}, the BIoT model and the adversary model are described. The optimal attacking strategy and the detection performance guarantee for the BIoT are investigated in Section \ref{Section_optimal}. Numerical simulations are  presented in Section \ref{Section_simulation}, and Section \ref{Section_conclusion} provides our conclusions.

\section{Blockchian-aided IoT Network And Adversary Models}\label{system_model}
\subsection{Blockchian-aided IoT Detection Network Model}\label{BIoT-DN model}

We consider a general BIoT model which is generalized from most existing works on BIoT applications \cite{zhuang2020blockchain,yang2021priscore,hjalmarsson2018blockchain,hassija2020traffic,lei2017blockchain, cha2018blockchain,lombardi2018blockchain,ling2019blockchain,li2018creditcoin,zhou2018beekeeper,yang2021privacy}.

Unlike the CIoT where IoT devices send their data to a cloud for further processing, the BIoT employs a blockchain which eliminates the need of a central authority to verify  and store IoT devices' data. According to functions, a BIoT  which performs a detection task is usually composed of three types of agents, that is, IoT devices (or "things"), miners, and decision makers. 


The first type of agent, the IoT device, forms the IoT layer of the BIoT. IoT devices are usually embedded with sensors, communication modules, software, and other technologies for the purpose of producing data and exchanging data with other agents over the BIoT. The second type of agent, the miner, forms the blockchain layer of the BIoT. The miners are assumed to have massive memory and computational power, and their duties are generating blocks which store the information produced by IoT devices.
Decision makers aim to make an accurate and consistent decision based on the information stored in their local copies of the blockchain. It is worth mentioning that in a BIoT, the agents owning local copies of the blockchain can play the role of a decision maker and make their own decisions.  Moreover, their decisions can be ensured to reach an agreement due to the blockchain consensus protocol. Hence,  a single agent may play different roles simultaneously in a BIoT. For example,  a miner can also play the role of a decision maker. In view of the popularity of the Proof-of-Work (PoW) consensus protocol, we assume that the BIoT considered in this paper employs the PoW consensus protocol. The BIoT employing other consensus protocols will be considered in future work. The working mechanism of the BIoT is similar to that of a PoW blockchain \cite{nakamoto2008bitcoin,antonopoulos2014mastering}, which are briefly summarized as follows.

\subsubsection{ Data Model} Let $N$ denote the number of IoT devices in the IoT layer, and each IoT device makes measurements under { the same} unknown binary hypothesis (i.e., $\mathcal{H}_0$ or $\mathcal{H}_1$) over time. 
Let $\bx_{j,l}$ denote the vector measurement made at the $j$-th IoT device in the $l$-th measurement sampling interval, $\forall l=1,2,...$ and $\forall j=1,2,...,N$. The $j$-th IoT device processes its raw measurement $\bx_{j,l}$ to produce a discrete data $u_{j,l}$ by employing a  function $\cQ_j(\cdot)$, that is, $u_{j,l} = \cQ_j(\bx_{j,l})\in \mathcal{K} \triangleq\{0,...,| \mathcal{K}|-1\}$ for each $j$ and $l$, where $ \mathcal{K}$ is the alphabet set of $u_{j,l}$ with cardinality equal to $| \mathcal{K}|$, and $u_{j,l}$ denotes the $j$-th IoT device's data produced in the $l$-th measurement sampling interval, $\forall j=1,2,...,N$.  It is worth mentioning that the process $\cQ_j(\cdot)$ implemented at each IoT device may be necessary in practice. For example, $\cQ_j(\cdot)$ can be an analog-to-digital converter employed at the $j$-th IoT device which is generally required for digital processing and digital communications.  We assume that  $\{u_{j,l}\}_{j,l}$ are statistically independent and identically distributed, and the alphabet sets of $u_{j,l}$ under $\mathcal{H}_1$ and $\mathcal{H}_0$ are the same. The probability mass functions  (PMFs) \cbk of $u_{j,l}$ are denoted by
 \begin{equation}{\bf{p}}^{H}=[p_0^{H},...,p_{|\mathcal{K}|-1}^{H}]^T\text{ and }{\bf{q}}^{H}=[q_0^{H},...,q_{|\mathcal{K}|-1}^{H}]^T,\end{equation}under hypotheses $\mathcal{H}_1$ and $\mathcal{H}_0$, respectively. As such, $\forall k=0,1,...,|\mathcal{K}|-1$,
\begin{equation}\label{equ_pq_H>0}
\begin{split}
\Pr\{u_{j,l}=k|\mathcal{H}_1\}=p_k^H>0\text{    and    }\Pr\{u_{j,l}=k|\mathcal{H}_0\}=q_k^H>0.
\end{split}
\end{equation} 
\subsubsection{ Data Exchange}
In each measurement sampling interval, every IoT device produces a data $u_{j,l}$, and the data $u_{j,l}$ will be sent along with its index $l$ to every miner. The communications over the BIoT are secured by asymmetric encryption which is similar to that of the PoW blockchain \cite{nakamoto2008bitcoin,antonopoulos2014mastering}. To be specific, each agent of a BIoT owns a public-private key pair that forms the digital identity of the agent. The public key is created from the private key and is available at the other agents, while the private key is only available at its owner. A secure hash algorithm (SHA), e.g., SHA-256 and SHA-512 \cite{pilkington2016blockchain}, is used in the data encryption process of every data exchange between two agents of the BIoT. Consider the data exchanges between the $j$-th IoT device and a miner as an example. 
In the $l$-th measurement sampling interval, the $j$-th IoT device firstly processes its message  that contains data $u_{j,l}$ and the data index $l$ by employing an SHA, and obtains a message digest. It then encrypts the message digest via its private key by using a digital signature algorithm, e.g., Elliptic Curve Digital Signature Algorithm \cite{dinh2018untangling}, and produces a digital signature. Finally, the $j$-th IoT device sends the data package consisting of the message and the corresponding digital signature to the miners of the BIoT. 

Once a miner receives the data package from an IoT device, it first decrypts the received digital signature via the public key of the IoT device, and obtains a message digest. Then, the miner processes the received message via the SHA to obtain another message digest. Only if these two message digests exactly match with each other, the authenticity of the received message from the IoT device is verified and the received message will be used in future processes. Otherwise, the received data package will be discarded and retransmission can take place. Note that the received digital signature at the miner can only be decrypted via the public key of the sender. It is worth mentioning that it is computationally intractable for an attacker to either find a different message which yields the same message digest or generate a valid digital signature for a fake message digest without the private key of the sender \cite{pilkington2016blockchain, dinh2018untangling}. Thus, the authenticity of the data package received at the miner and the identity of the sender can be validated and secured,  which can prevent impersonation of the sender.

\subsubsection{Block Mining and Consensus Protocol}

Each miner maintains a local copy of the blockchain which is a chronologically ordered sequence of cryptographically linked blocks. For each $l$, after collecting and verifying the authenticity of the messages $\{u_{j,l}\}_{\forall j,l}$ \cbk from all the IoT devices, each miner 
 constructs a local block which is a candidate for the $l$-th block of the blockchain. To be specific, each miner puts $\{u_{j,l}\}_{\forall j,l}$, \cbk the data index $l$, and the digital signatures associated with the IoT device messages into a block with a header. The header of a block consists of a discrete timestamp, a difficulty value, the hash value of the last block (parent block) of the longest branch in the miner's local copy of the blockchain,  a Merkle Root which is the root of the Merkle tree constructed by recursively hashing pairs of data packages until there is only one hash \cite{nakamoto2008bitcoin,antonopoulos2014mastering}, and a number called  nonce which is the solution to a puzzle problem. The hash value of the parent block in  the header cryptographically links the block to its parent block. 

Next, the miners compete with each other in solving a difficult PoW puzzle for their local blocks, which is called mining. The PoW puzzle is to find a nonce for a block such that the hash value of the block is smaller than a given value, i.e., the Difficulty in the header \cite{nakamoto2008bitcoin,antonopoulos2014mastering}. Each miner searches for such a nonce for its local block via brute-force search. Once a miner solves its PoW puzzle, i.e., find a valid nonce value for its local block which meets the hash value requirement, it has successfully mined its local block, and it broadcasts its local block to all the other miners and the decision makers. After the other miners and the decision makers complete the verification of the authenticity of the data in the received block and the verification that the hash value of the received block is indeed smaller than the difficulty value and the data index stored in the received block is just one greater than that in its parent block, they add this block after its parent block in their local copies of the blockchain, and switch to work on mining the next block. 
The block mining process described above can be considered as a hashing competition among the miners, where the probability that a miner solves its PoW puzzle first among the miners is proportional to its hash rate which is defined as the ratio of its computational power to the total computational power in the network \cite{nakamoto2008bitcoin}.
We assume that every decision maker has enough memory resources to store a full copy of the blockchain, and each decision maker only employs the IoT devices' data stored in the longest branch of its local copy of the blockchain to make its decision between two hypotheses. 
Note that for detection problems, it is generally true that the more the data, the smaller the decision error \cite{poor2013introduction}. For a detection task, each decision maker only makes its decision when the number of blocks in the longest branch of its local copy of the blockchain grows to $L$ so that the decision error can be guaranteed to meet some prescribed requirement. As such,  from the perspective of this detection task, the growth of the blockchain can be considered being terminated once the longest branch of the blockchain employed in the BIoT grows to $L$ blocks.

\subsection{Adversary Model}\label{Section_Adversarymodel}
The vulnerability of  a BIoT comes from both the vulnerability of its IoT devices and that of the blockchain employed in the BIoT, which provides adversaries an opportunity to undermine the detection performance of the BIoT. In particular, with the goal of misleading the decision makers of the BIoT into making erroneous decisions, adversaries can 
jointly exploit both the vulnerability of IoT devices and  that of the blockchain employed in the BIoT to falsify the data utilized by the decision makers.\cbk
\subsubsection{Attacks against IoT Devices} In order to falsify IoT devices' data in the blockchain without being perceived in the verification and validation processes described in Section \ref{BIoT-DN model}, an attacker first has to hack into IoT devices and obtain their private keys to generate valid falsified data packages\footnote {Valid falsified data packages mean those which can pass the cryptographic algorithms based authenticity verification process employed by the BIoT.}. If an IoT device has been hacked and controlled by an attacker, the attacker can use its  private key to generate valid falsified data packages  and send them to  miners. Once the blocks which contain valid falsified data packages are successfully mined, the valid falsified data packages will be stored in the decision makers' local copies of the blockchain without being perceived.

Generally, cybersecurity measures are taken at IoT devices to keep attackers from hacking into the IoT devices and stealing their private keys. To this end, an attacker may not be able to hack into every IoT device within a limited time period. But the longer the period that the attacker spends on hacking into an IoT device, the higher the probability that the attacker successfully hacks into the target IoT device. For example, as a basic preventative measure, IoT devices can be equipped with password protection to prevent hacking. In consequence, an attacker has to employ a brute force attack which works through all possible keystrokes hoping to guess the password correctly. The longer the hacking period, the higher the probability that the attacker guesses the password correctly \cite{stiawan2019investigating}.

\subsubsection{Attacks against Blockchain} As pointed out in \cite{nakamoto2008bitcoin,zaghloul2020bitcoin,antonopoulos2014mastering}, if an attacker aims to falsify the data which have already been stored in the blockchain of a BIoT, it has to  launch a successful double-spending attack\footnote{Double-spending attacks were firstly introduced in financial blockchain applications \cite{nakamoto2008bitcoin,antonopoulos2014mastering,karame2015misbehavior}. That we use the same terminology is because the attacking mechanism here is the same as that of  double-spending attacks against blockchain financial applications.} on the blockchain. To be specific, the attacker has to generate valid counterfeit blocks\footnote{If a block contains any falsified data package, this block is referred to as a counterfeit block. Otherwise, it is called an authentic block.} containing valid falsified data packages (i.e., solve PoW puzzles for the valid counterfeit blocks) to form a counterfeit branch in the blockchain which must surpass the authentic branch.
A DSA is deemed successful if its counterfeit branch grows to $L$ blocks before the authentic branch does, and hence its counterfeit branch is the longest branch in the blockchain when decision makers make their decisions. Under a successful DSA, the decisions will be made based on the falsified data stored in a counterfeit branch, and therefore, can be seriously misled.

In this paper, we consider an adversary model which jointly exploits the vulnerability of IoT devices and that of the blockchain employed in the BIoT. The IoT devices usually stay in sleeping mode during long intervals of inactivity to reduce energy consumption and reduce the risk of being targeted by an attacker \cite{neshenko2019demystifying,kaur2015energy,ahmed2022energy}. Once the BIoT starts to work, its IoT devices wake up to make measurements of an unknown hypothesis and connect to the Internet to transfer their data to the miners in the network, which provides attackers an opportunity to hack into these IoT devices. We assume that there is a portion of miners, called malicious miners, which are under the command of an attacker. The attacker first commands the malicious miners to attempt to hack into every IoT device via the Internet, while the honest miners, which are not controlled by the attacker, follow the standard blockchain protocol to mine blocks for storing IoT devices' data, which form an authentic branch in the blockchain.  
At a time instant $t_0$, we assume that the probability that an IoT device has been hacked and controlled by the attacker is $\alpha$ ($\alpha \ge 0$), and the honest miners have built $L_0$ ($L_0 \ge 0$)  blocks in the authentic branch of the blockchain. Thus, the expected percentage of IoT devices which are hacked and controlled by the attacker is $\alpha$ at $t_0$. If an IoT device has been hacked by the attacker, then the IoT device is called a malicious IoT device. Otherwise, it is called an honest IoT device. Starting from the time instant $t_0$, the malicious IoT devices' data can be deliberately falsified by the attacker since the attacker has already controlled these IoT devices and obtained their private keys to generate valid falsified data packages. As a result, the blocks of the authentic branch of the blockchain whose indices are greater than
$L_0$ are counterfeit blocks since in these blocks, the data from the malicious IoT devices are falsified by the attacker without the need to launch a DSA.  Next, in order to falsify the malicious IoT devices' data which have already been stored in the authentic branch's blocks whose indices are smaller than or equal to $L_0$, we assume that the attacker devotes all the computational power of the malicious miners to launching a DSA on the blockchain. Let $\mathcal{A}$ denote the set of malicious IoT devices, and we assume that the attacker aims at falsifying the  malicious IoT devices' data $\{u_{j,l}\}_{j\in\mathcal{A}, L_A\le l\le L_0}$, which have been stored in the authentic branch's blocks with indices greater than or equal to $L_A$, to $\{\tilde{u}_{j,l}\}_{j\in\mathcal{A}, L_A\le l\le L_0}$ where\footnote{If $L_0=0$, then  the attacker can falsify all the malicious IoT devices' data stored in the blockchain without the need to launch a DSA. Therefore, when $L_0=0$, we define $L_A=0$ which signifies that the attacker does not launch a DSA. For the cases where $L_0>0$, $L_A$ is greater than $0$.} $0\le L_A\le L_0$.
For the cases where $L_0>0$, the counterfeit branch built by the malicious miners diverges from the authentic branch at the ($L_A-1$)-th block of the authentic branch.
It is worth mentioning that the counterfeit branch includes the authentic blocks with indices smaller than $L_A$ and the counterfeit blocks in the counterfeit branch 
whose indices are greater than or equal to $L_A$. The authentic branch consists of the authentic blocks with indices from $1$ to $L_0$ and the counterfeit blocks in the authentic branch 
whose indices are greater than or equal to $(L_0+1)$. Once the attacker extends the counterfeit branch to become longer than the authentic branch, all the honest miners switch from extending the authentic branch to working on extending the counterfeit branch according to the PoW consensus protocol \cite{nakamoto2008bitcoin,antonopoulos2014mastering,zaghloul2020bitcoin,karame2015misbehavior}. Therefore, the authentic branch stops growing, and the counterfeit branch will remain the longest branch in the blockchain as time goes by which implies that a successful DSA has been launched.


It is worth mentioning that the honest IoT devices' data in counterfeit blocks are authentic and cannot be falsified by the attacker since the attacker does not have their private keys to generate valid falsified data packages for the honest IoT devices.  In addition, the attacker has to command the malicious miners to find a valid nonce value for each counterfeit block in the counterfeit branch so that the counterfeit block can be accepted by the decision makers and the other miners in their local copies of the blockchain due to the PoW consensus protocol.  If the counterfeit branch built by the attacker is the longest branch in the blockchain when the decision makers make their decisions, the counterfeit branch is considered as the only valid branch by the decision makers according to the PoW consensus protocol. In consequence, the attacker may be able to fool the decision makers into reaching erroneous decisions since every decision maker employs the falsified IoT devices' data stored in the counterfeit branch to make its decision.  

We assume that if the $j$-th IoT device is controlled by the attacker, the attacker falsifies the $j$-th IoT device's data $\{u_{j,l}\}_{l=L_A}^L$ to $\{\tilde{u}_{j,l}\}_{l=L_A}^L$ which is an independent and identically distributed sequence and follows the malicious-data  PMFs \cbk
\begin{equation}\label{equ_qb}{\bf{p}}^{B}=[p_0^{B},...,p_{|\mathcal{K}|-1}^{B}]^T\text{ and }{\bf{q}}^{B}=[q_0^{B},...,q_{|\mathcal{K}|-1}^{B}]^T,\end{equation}under hypotheses $\mathcal{H}_1$ and $\mathcal{H}_0$, respectively. As such, if the $j$-th IoT device is malicious, then $\forall l=L_A,L_A+1,...,L $ and $\forall k=0,1,...,| \mathcal{K}|-1$,
\begin{equation}\label{equ_qb1}
\Pr\{\tilde{u}_{j,l}=k|\mathcal{H}_1\}=p_k^B  \text{  and  }  \Pr\{\tilde{u}_{j,l}=k|\mathcal{H}_0\}=q_k^B.
\end{equation}
It is worth mentioning that if $L_0=L$, then when the attacker starts to launch a DSA, the authentic branch has already grown to $L$ blocks, and each decision maker has already made its decision. Therefore, it is pointless for the attacker to launch a DSA to impair the detection performance of the BIoT, and hence the case that $L_0=L$ is a trivial case. The case where $\alpha=0$ is also a trivial case since if $\alpha=0$, the attacker cannot falsify any data stored in the blockchain since the attacker does not obtain any IoT device's private key to generate valid falsified data. Before proceeding, to avoid these trivial cases, we make the following assumption throughout this paper.
\begin{assumption} \label{Assumption_trivial2}
We assume that $0<\alpha\le 1$ and $L_A\le L_0<L$.
\end{assumption}

\section{Optimal Attacking Strategy and Guaranteed Detection Performance}\label{Section_optimal}
In this section, we investigate the detection performance of a BIoT in the presence of attacks, and pursue a guaranteed detection performance under any attacks.
It can be shown that the probability that an attacker successfully launches a DSA is a constant depending on $L_0$, $L_A$, $L$, and the malicious miners' hash rate \cite{jiang2023vulnerability}. 
We use $P_s$ to denote the probability that the attacker successfully launches a DSA on the BIoT.

When the longest branch of the blockchain grows to $L$ blocks, let $\hat{\textbf{u}}_j\triangleq[\hat{u}_{j,1},\hat{u}_{j,2},...,\hat{u}_{j,L}]^T$ and $\textbf{U}\triangleq[\hat{\textbf{u}}_1,\hat{\textbf{u}}_2,...,\hat{\textbf{u}}_N]$ denote the data of the $j$-th IoT device and all IoT devices' data stored in the longest branch of the blockchain, respectively. Note that $\hat u _{j,l} = u_{j,l}$, $\forall l \in \{1,2,...,L_A-1\}$, while\footnote{ We define $L_A^+\triangleq\max\{L_A,1\}$ to subsume the case where $L_A=0$ and $L_0=0$.} $\forall l \in \{L_A^+, L_A^++1,...,L \}$, $\hat u _{j,l}$ can be either $u_{j,l}$ or $\tilde{u}_{j,l}$ depending on whether the $j$-th IoT device is malicious and whether the DSA launched by the attacker is successful. 
Define $\textbf{R}_i\triangleq[\textbf{r}_{i}^{(1)},\textbf{r}_{i}^{(2)},...,\textbf{r}_{i}^{(N)}]\in\mathcal{R}$ where $\textbf{r}_{i}^{(j)}\triangleq[r^{(j)}_{i,1},r^{(j)}_{i,2},...,r^{(j)}_{i,L}]^T\in\mathcal{O}$, $\forall j\in\{1,2,...,N\}$, and $r^{(j)}_{i,l}\in\mathcal{K}$ for any $i$ and $l$. $\mathcal{R}$ denotes the set of all possible $\textbf{R}_i$ with cardinality $|\mathcal{R}|=|{\cO}|^N$, and  $\cal O$ denotes the set of all possible $\textbf{r}_{i}^{(j)}$ with cardinality $|{\cal O}|=|{\cal K}|^L$. Under hypothesis $\mathcal{H}_1$, for any $\textbf{R}_i\in\mathcal{R},$
\begin{align}\notag
\Pr&\left\{\textbf{U}=\textbf{R}_i|\mathcal{H}_1\right\}\\ \notag
=&\Pr\left\{\textbf{U}=\textbf{R}_i|\mathcal{E},\mathcal{H}_1\right\}\times\Pr\left\{\mathcal{E}|\mathcal{H}_1\right\}\\  \notag
&+\Pr\left\{\textbf{U}=\textbf{R}_i\left|\mathcal{E}^{\rm C},\mathcal{H}_1\right.\right\}\times\Pr\left\{\left.\mathcal{E}^{\rm C}\right|\mathcal{H}_1\right\}\\ \notag
=&P_s\left(\prod_{j=1}^N\Pr\left\{\left.\hat{\textbf{u}}_j=\textbf{r}_{i}^{(j)}\right|\mathcal{E},\mathcal{H}_1\right\}\right)\\
&+(1-P_s)\left(\prod_{j=1}^N\Pr\left\{\left.\hat{\textbf{u}}_j=\textbf{r}_{i}^{(j)}\right|\mathcal{E}^{\rm C},\mathcal{H}_1\right\}\right),\label{DM_under_H1_explain}
\end{align}
due to the fact that $\{\hat{\textbf{u}}_j\}_{j=1}^N$ are statistically independent. The notation $\mathcal{E}$ in (\ref{DM_under_H1_explain}) stands for the event that the DSA is successful, and $\mathcal{E}^{\rm C}$ is the event that the DSA is not successful.

If the $j$-th IoT device is honest, then $\hat{u}_{j,l}=u_{j,l},\forall l\in \{1,2,...,L\}$.
On the other hand, if the $j$-th IoT device is malicious, then $\hat{u}_{j,l}=u_{j,l},\forall l\in\{1,2,...,L_A-1\}$ and $\hat{u}_{j,l}=\tilde{u}_{j,l}$, $\forall l\in \{L_0,L_0+1,...,L\}$. Moreover, if the DSA launched by the attacker is successful, i.e., $\mathcal{E}$ happens, then
 $\hat{u}_{j,l}=\tilde{u}_{j,l}$, $\forall l\in\{L_A^+, L_A^++1,...,L_0-1\}$. Otherwise, $\hat{u}_{j,l}=u_{j,l}$, $\forall l\in\{L_A^+, L_A^++1,...,L_0-1\}$.
As such, we can obtain\footnote{We define $\prod_{l_1}^{l_2}(\cdot)=1$ if $l_2<l_1$.}
\begin{align}\label{DSA_S_under_H1_1} \notag
\Pr&\left\{\left.\hat{\textbf{u}}_j=\textbf{r}_{i}^{(j)}\right|\mathcal{E},\mathcal{H}_1\right\}\\ \notag
=&\Pr\left\{\left.\hat{\textbf{u}}_j=\textbf{r}_{i}^{(j)}\right|j\not\in\mathcal{A},\mathcal{E},\mathcal{H}_1\right\}\\ \notag
&\times\Pr\left\{\left.j\not\in\mathcal{A}\right|\mathcal{E},\mathcal{H}_1\right\}\\ \notag
&+\Pr\left\{\left.\hat{\textbf{u}}_j=\textbf{r}_{i}^{(j)}\right|j\in\mathcal{A},\mathcal{E},\mathcal{H}_1\right\}\\ \notag
&\times\Pr\left\{\left.j\in\mathcal{A}\right|\mathcal{E},\mathcal{H}_1\right\}
\\=&(1-\alpha)\prod_{l=1}^Lp^H_{r^{(j)}_{i,l}}+\alpha\prod_{l=1}^{L_A-1}p^H_{r^{(j)}_{i,l}}\prod_{l=L_A^+}^Lp^B_{r^{(j)}_{i,l}},
\end{align} due to the fact that $\{\hat{u}_{j,l}\}_l$ are statistically independent,
and similar to (\ref{DSA_S_under_H1_1}), we can obtain
\begin{align}\notag
\Pr&\left\{\left.\hat{\textbf{u}}_j=\textbf{r}_{i}^{(j)}\right|\mathcal{E}^{\rm C},\mathcal{H}_1\right\}\\ 
=&(1-\alpha)\prod_{l=1}^Lp^H_{r^{(j)}_{i,l}}+\alpha\prod_{l=1}^{L_0}p^H_{r^{(j)}_{i,l}}\prod_{l=L_0+1}^Lp^B_{r^{(j)}_{i,l}}.
\label{DSA_F_under_H1_1}
\end{align}
From (\ref{DM_under_H1_explain}), (\ref{DSA_S_under_H1_1}), and (\ref{DSA_F_under_H1_1}), we can obtain that under  $\cH_1$, for any $\textbf{R}_i\in\mathcal{R}$,
\begin{align}\notag
\Pr&\{\textbf{U}=\textbf{R}_i|\mathcal{H}_1\}\\ \notag
=&P_s\left\{\prod_{j=1}^N\left[(1-\alpha)\prod_{l=1}^Lp^H_{r^{(j)}_{i,l}}+\alpha\prod_{l=1}^{L_A-1}p^H_{r^{(j)}_{i,l}}\prod_{l=L_A^+}^Lp^B_{r^{(j)}_{i,l}}\right]\right\}
\end{align}
\begin{align} \notag
\ &+(1-P_s)\Bigg\{\prod_{j=1}^N\left[(1-\alpha)\prod_{l=1}^Lp^H_{r^{(j)}_{i,l}}+\alpha\prod_{l=1}^{L_0}p^H_{r^{(j)}_{i,l}}\right.\\ 
&\left.\times\prod_{l=L_0+1}^Lp^B_{r^{(j)}_{i,l}}\right]\Bigg\}.\label{DM_under_H1}
\end{align}
Similar to (\ref{DM_under_H1}),  we also can obtain that under hypothesis $\cH_0$, for any $\textbf{R}_i\in\mathcal{R}$,
\begin{align}\notag
\Pr&\{\textbf{U}=\textbf{R}_i|\mathcal{H}_0\}\\ \notag
=&P_s\left\{\prod_{j=1}^N\left[(1-\alpha)\prod_{l=1}^Lq^H_{r^{(j)}_{i,l}}+\alpha\prod_{l=1}^{L_A-1}q^H_{r^{(j)}_{i,l}}\prod_{l=L_A^+}^Lq^B_{r^{(j)}_{i,l}}\right]\right\}\\ \notag
&+(1-P_s)\Bigg\{\prod_{j=1}^N\left[(1-\alpha)\prod_{l=1}^Lq^H_{r^{(j)}_{i,l}}+\alpha\prod_{l=1}^{L_0}q^H_{r^{(j)}_{i,l}}\right.\\ 
&\left.\times\prod_{l=L_0+1}^Lq^B_{r^{(j)}_{i,l}}\right]\Bigg\}.\label{DM_under_H0}
\end{align}
By evaluating (\ref{DM_under_H1}) and (\ref{DM_under_H0}) for all possible $\textbf{R}_i$, we can obtain the probability mass functions $\textbf{p}\triangleq[p_1,p_2,...,p_{|\mathcal{R}|}]^T$ and $\textbf{q}\triangleq[q_1,q_2,...,q_{|\mathcal{R}|}]^T$ of $\textbf{U}$ under hypotheses $\cH_1$ and $\cH_0$, respectively, where $p_i\triangleq \Pr\{\textbf{U}=\textbf{R}_i|\mathcal{H}_1\}$ and $q_i\triangleq \Pr\{\textbf{U}=\textbf{R}_i|\mathcal{H}_0\}$ are the probabilities of the event $\{\textbf{U}=\textbf{R}_i\}$ under hypotheses $\cH_1$ and $\cH_0$, respectively.

According to Stein's lemma, the KLD indicates the best error exponent of the miss probability under the Neyman-Pearson setup  \cite{Cover:1991}. To this end, we choose the KLD as the performance metric for the BIoT, and pursue a guaranteed detection performance of the BIoT for all possible  malicious-data distributions $\textbf{p}^B$ and $\textbf{q}^B$.

It is seen from (\ref{DM_under_H1}) and (\ref{DM_under_H0}) that the probability mass functions $\textbf{p}$ and $\textbf{q}$ are functions of $\textbf{p}^B$ and $\textbf{q}^B$ in (\ref{equ_qb}), and hence, the  KLD $D(\textbf{q}||\textbf{p})$ between $\textbf{q}$ and $\textbf{p}$ is also a function of $\textbf{p}^B$ and $\textbf{q}^B$, which can be represented by $D_0(\textbf{p}^B,\textbf{q}^B)$, that is,
\begin{align}\label{D0_function}\notag
D_0\left(\textbf{p}^B,\textbf{q}^B\right) \triangleq&  D(\textbf{q}||\textbf{p}) \\
=& \sum_{i=1}^{|\mathcal{R}|}\Pr\{\textbf{U}=\textbf{R}_i|\mathcal{H}_0\}\log\frac{\Pr\{\textbf{U}=\textbf{R}_i|\mathcal{H}_0\}}{\Pr\{\textbf{U}=\textbf{R}_i|\mathcal{H}_1\}}.
\end{align}
A guaranteed KLD  for all possible  malicious-data distributions can be obtained by solving the following optimization problem
\begin{subequations} \label{original_KLD}
\begin{align} \label{original_KLD_obj}
\mathop{\min}\limits_{\textbf{p}^B,\textbf{q}^B} \; &D_0\left(\textbf{p}^B,\textbf{q}^B\right)\\  
\text{s.t.} \quad &0\le p_k^B\le 1,0\le q_k^B\le 1, \; \forall k\in\mathcal{K},\\ \label{original_KLD_constraint}
&\sum_{k\in\mathcal{K}}p_k^B=1,\sum_{k\in\mathcal{K}}q_k^B=1.
\end{align}
\end{subequations}
In general, the objective function  $D_0(\textbf{p}^B,\textbf{q}^B)$  in (\ref{original_KLD_obj}) is a nonconvex function of $\textbf{p}^B$ and $\textbf{q}^B$. Thus, it is generally intractable to obtain the optimal solution to (\ref{original_KLD}). 
To address this issue, we propose a relaxation method which converts the nonconvex optimization problem in (\ref{original_KLD}) into a convex optimization problem, which is elaborated below.

%


Note that for each $\textbf{R}_i\in\mathcal{R}$, $\Pr\{\textbf{U}=\textbf{R}_i|\mathcal{H}_1\}$ in (\ref{DM_under_H1}) can be rewritten as
\begin{align}\notag
\Pr&\{\textbf{U}=\textbf{R}_i|\mathcal{H}_1\}\\ \notag
=&\underbrace{(1-\alpha)^N\prod_{l_1=1}^Lp^H_{r^{(1)}_{i,l_1}}\prod_{l_2=1}^Lp^H_{r^{(2)}_{i,l_2}}\cdots\prod_{l_N=1}^Lp^H_{r^{(N)}_{i,l_N}}}_\text{$\triangleq\Psi_i$}\\
&+\sum_{m\in\mathcal{M}}\Gamma_{i,m}{P}^{(1)}_{i,m}
+\sum_{m\in\mathcal{M}}\Lambda_{i,m}{P}^{(2)}_{i,m},\label{DM_under_H1_expand}
\end{align}
where $\Psi_i$ is the probability that $\textbf{U}=\textbf{R}_i$ and all the $N$ IoT devices are honest under $\mathcal{H}_1$. $\mathcal{M}\triangleq\{1,2,...,2^N-1\}$ whose elements are used to specify all possible states of the $N$ IoT devices except for the state that all the $N$ IoT devices are honest. The state of the $j$-th IoT device is indicated by $S_m^N(j)$ which is the $j$-th digit of the $N$-bit-long binary sequence  $\textbf{S}_m^N$ converted from a decimal number $m$. We define that  $S_m^N(j)=1$ indicates that the $j$-th IoT device is malicious, and $S_m^N(j)=0$ indicates that the $j$-th IoT device is honest. In (\ref{DM_under_H1_expand}), $\Gamma_{i,m}$, ${P}^{(1)}_{i,m}$, $\Lambda_{i,m}$, and ${P}^{(2)}_{i,m}$ are defined as\footnote{For simplicity of notation, we define $0^0=1$, and hence when $\alpha=1$ and $n_m=N$ (i.e., $m=2^N-1$), we have $(1-\alpha)^{N-n_m}=1$.} 
\begin{align}\label{def_Gamma}&\Gamma_{i,m}\triangleq P_s(1-\alpha)^{N-n_m}\alpha^{n_m}\prod_{j=1}^N\gamma_{i,m}^{(j)},\\
	&\label{def_P1}{P}^{(1)}_{i,m}\triangleq \prod_{j=1}^N{p}^{(1)}_{i,j,m},\\
\label{def_Lambda}&\Lambda_{i,m}\triangleq (1-P_s)(1-\alpha)^{N-n_m}\alpha^{n_m}\prod_{j=1}^N\lambda_{i,m}^{(j)},\\
&\label{def_P2}{P}^{(2)}_{i,m}\triangleq \prod_{j=1}^N{p}^{(2)}_{i,j,m},
\\ \label{equ_gamma}
&\gamma_{i,m}^{(j)}\triangleq \left\{\begin{array}{l}
\prod_{l=1}^{L_A-1}p^H_{r^{(j)}_{i,l}}\text{, if }S^N_m(j)=1,\\
\prod_{l=1}^{L}p^H_{r^{(j)}_{i,l}}\text{, if }S^N_m(j)=0,
\end{array}\right.
\\\label{equ_p1}
&{p}^{(1)}_{i,j,m}\triangleq \left\{\begin{array}{l}
\prod_{l=L_A^+}^{L}p^B_{r^{(j)}_{i,l}}\text{, if }S^N_m(j)=1,\\
1\text{, if }S^N_m(j)=0,
\end{array}\right.
\\\label{equ_lambda}
&\lambda_{i,m}^{(j)}\triangleq \left\{\begin{array}{l}
\prod_{l=1}^{L_0}p^H_{r^{(j)}_{i,l}}\text{, if }S^N_m(j)=1,\\
\prod_{l=1}^{L}p^H_{r^{(j)}_{i,l}}\text{, if }S^N_m(j)=0,
\end{array}\right.
\\\label{equ_p2}
&{p}^{(2)}_{i,j,m}\triangleq \left\{\begin{array}{l}
\prod_{l=L_0+1}^{L}p^B_{r^{(j)}_{i,l}}\text{, if }S^N_m(j)=1,\\
1\text{, if }S^N_m(j)=0,
\end{array}\right.
\end{align}
and $n_m$ is the number of malicious IoT devices when the states of the $N$ IoT devices are specified by $\textbf{S}_m^N$. In other words,
\begin{align}\label{equ_nm}n_m=\sum_{j=1}^N\mathbb{1}_{\{1\}}\left[S^N_m(j)\right],\end{align}
where $\mathbb{1}_{\mathcal{X}}[x]=1$, if $x\in \mathcal{X}$, and $\mathbb{1}_{\mathcal{X}}[x]=0$, if $x\not\in \mathcal{X}$. For example, if $N=3$ and $m=1$, then $\textbf{S}^3_1=001$, $S^3_1(1)=0$, $S^3_1(3)=1$, and $n_1=1$. As such, $\Gamma_{i,m}{P}^{(1)}_{i,m}$ is the probability that $\textbf{U}=\textbf{R}_i$,  the DSA is successful, and the states of the $N$ IoT devices are specified by $\textbf{S}_m^N$ under $\mathcal{H}_1$. $\Lambda_{i,m}{P}^{(2)}_{i,m}$ is the probability that $\textbf{U}=\textbf{R}_i$,  the DSA is not successful, and the states of the $N$ IoT devices are specified by $\textbf{S}_m^N$ under $\mathcal{H}_1$. 

Similar to (\ref{DM_under_H1_expand}), $\Pr\{\textbf{U}=\textbf{R}_i|\mathcal{H}_0\}$ in  (\ref{DM_under_H0}) can be rewritten as
\begin{align}\notag
\Pr&\{\textbf{U}=\textbf{R}_i|\mathcal{H}_0\}\\ \notag
=&\underbrace{(1-\alpha)^N\prod_{l_1=1}^Lq^H_{r^{(1)}_{i,l_1}}\prod_{l_2=1}^Lq^H_{r^{(2)}_{i,l_2}}\cdots\prod_{l_N=1}^Lq^H_{r^{(N)}_{i,l_N}}}_\text{$\triangleq\Theta_i$}\\
&+\sum_{m\in\mathcal{M}}\Delta_{i,m}{Q}^{(1)}_{i,m}
+\sum_{m\in\mathcal{M}}\Phi_{i,m}{Q}^{(2)}_{i,m},\label{DM_under_H0_expand}
\end{align}
where 
\begin{align}\label{def_Delta}&\Delta_{i,m}\triangleq P_s(1-\alpha)^{N-n_m}\alpha^{n_m}\prod_{j=1}^N\delta_{i,m}^{(j)},\\
&\label{def_Q1}{Q}^{(1)}_{i,m}\triangleq \prod_{j=1}^N{q}^{(1)}_{i,j,m},\\
\label{def_Phi}&\Phi_{i,m}\triangleq (1-P_s)(1-\alpha)^{N-n_m}\alpha^{n_m}\prod_{j=1}^N\phi_{i,m}^{(j)},\\
&\label{def_Q2}{Q}^{(2)}_{i,m}\triangleq \prod_{j=1}^N{q}^{(2)}_{i,j,m},
\\\label{equ_delta}
&\delta_{i,m}^{(j)}\triangleq \left\{\begin{array}{l}
\prod_{l=1}^{L_A-1}q^H_{r^{(j)}_{i,l}}\text{, if }S^N_m(j)=1,\\
\prod_{l=1}^{L}q^H_{r^{(j)}_{i,l}}\text{, if }S^N_m(j)=0,
\end{array}\right.
\\\label{equ_q1}
&{q}^{(1)}_{i,j,m}\triangleq \left\{\begin{array}{l}
\prod_{l=L_A^+}^{L}q^B_{r^{(j)}_{i,l}}\text{, if }S^N_m(j)=1,\\
1\text{, if }S^N_m(j)=0,
\end{array}\right.
\\\label{equ_phi}
&\phi_{i,m}^{(j)}\triangleq \left\{\begin{array}{l}
\prod_{l=1}^{L_0}q^H_{r^{(j)}_{i,l}}\text{, if }S^N_m(j)=1,\\
\prod_{l=1}^{L}q^H_{r^{(j)}_{i,l}}\text{, if }S^N_m(j)=0,
\end{array}\right.
\\\label{equ_q2}
&{q}^{(2)}_{i,j,m}\triangleq \left\{\begin{array}{l}
\prod_{l=L_0+1}^{L}q^B_{r^{(j)}_{i,l}}\text{, if }S^N_m(j)=1,\\
1\text{, if }S^N_m(j)=0.
\end{array}\right.
\end{align}
In fact, $\Delta_{i,m}{Q}^{(1)}_{i,m}$ is the probability that $\textbf{U}=\textbf{R}_i$, the DSA is successful, and the states of the $N$ IoT devices are indicated by $\textbf{S}_m^N$ under $\mathcal{H}_0$. $\Phi_{i,m}{Q}^{(2)}_{i,m}$ is the probability that $\textbf{U}=\textbf{R}_i$, the DSA is not successful, and the states of the $N$ IoT devices are specified by $\textbf{S}_m^N$ under $\mathcal{H}_0$.
Note that from their definitions, ${P}_{i,m}^{(1)}$, ${P}_{i,m}^{(2)}$, ${Q}_{i,m}^{(1)}$,  and ${Q}_{i,m}^{(2)}$ are products of quantities between zero and one. Hence we have ${P}_{i,m}^{(1)},{P}_{i,m}^{(2)},{Q}_{i,m}^{(1)},{Q}_{i,m}^{(2)}\in[0,1]$ for any $i=1,2,...,|\mathcal{R}|$ and $m\in\mathcal{M}$.

For simplicity of notation, we define $\mathcal{I}\triangleq\{1,2,...,|\mathcal{R}|\}$. Note from (\ref{equ_pq_H>0}), (\ref{def_Gamma}), (\ref{def_Lambda}), (\ref{equ_gamma}), (\ref{equ_lambda}), (\ref{def_Delta}), (\ref{def_Phi}), (\ref{equ_delta}), and (\ref{equ_phi}) that for different $i\in\mathcal{I}$, the signs of $\Gamma_{i,m}$, $\Lambda_{i,m}$, $\Delta_{i,m}$, and $\Phi_{i,m}$ remain unchanged, respectively. Hence, we can define $\mathcal{M}_{\Gamma}\triangleq\{m\in\mathcal{M}|\Gamma_{i,m}\not=0\}$, $\mathcal{M}_{\Lambda}\triangleq\{m\in\mathcal{M}|\Lambda_{i,m}\not=0\}$, $\mathcal{M}_{\Delta}\triangleq\{m\in\mathcal{M}|\Delta_{i,m}\not=0\}$, and $\mathcal{M}_{\Phi}\triangleq\{m\in\mathcal{M}|\Phi_{i,m}\not=0\}$ for any $i\in\mathcal{I}$. We also define $\textbf{P}_m^{(1)}$, $\forall m\in\mathcal{M}_{\Gamma}$, $\textbf{P}_m^{(2)}$, $\forall m\in\mathcal{M}_{\Lambda}$, $\textbf{Q}_m^{(1)}$, $\forall m\in\mathcal{M}_{\Delta}$, and  $\textbf{Q}_m^{(2)}$, $\forall m\in\mathcal{M}_{\Phi}$ as vectors stacking  $\{{P}_{i,m}^{(1)}\}_{i\in\mathcal{I}}$, $\{{P}_{i,m}^{(2)}\}_{i\in\mathcal{I}}$, $\{{Q}_{i,m}^{(1)}\}_{i\in\mathcal{I}}$, and $\{{Q}_{i,m}^{(2)}\}_{i\in\mathcal{I}}$, respectively. Note from (\ref{DM_under_H1_expand}) that if $\Gamma_{i,m}=0$, then $P_{i,m}^{(1)}$ does not affect $\Pr\{\textbf{U}=\textbf{R}_i|\mathcal{H}_1\}$, and thus does not affect $D_0(\textbf{p}^B,\textbf{q}^B)$. Similarly, if  either $\Lambda_{i,m}=0$, $\Delta_{i,m}=0$, or $\Phi_{i,m}=0$, then the corresponding $P_{i,m}^{(2)}$, $Q_{i,m}^{(1)}$, or $Q_{i,m}^{(2)}$ does not affect $D_0(\textbf{p}^B,\textbf{q}^B)$. In light of this, we define $\boldsymbol{\theta}$ as a one by $[\sum_{i\in\mathcal{I}}(|\mathcal{M}_{\Gamma}|+|\mathcal{M}_{\Lambda}|+|\mathcal{M}_{\Delta}|+|\mathcal{M}_{\Phi}|)]$ vector stacking $\{{P}_{i,m}^{(1)}\}_{i\in\mathcal{I},m\in\mathcal{M}_{\Gamma}}$, $\{{{P}}_{i,m}^{(2)}\}_{ i\in\mathcal{I},m\in\mathcal{M}_{\Lambda}}$, $\{{{Q}}_{i,m}^{(1)}\}_{ i\in\mathcal{I},m\in\mathcal{M}_{\Delta}}$, and $\{{{Q}}_{i,m}^{(2)}\}_{i\in\mathcal{I}, m\in\mathcal{M}_{\Phi}}$ which are the parameters of $D_0(\textbf{p}^B,\textbf{q}^B)$. 
It is seen from (\ref{DM_under_H1}), (\ref{DM_under_H0}), (\ref{D0_function}), (\ref{DM_under_H1_expand}), and (\ref{DM_under_H0_expand}) that $D_0(\textbf{p}^B,\textbf{q}^B)$ can be rewritten as
\begin{align} \notag
&D_1\left(\boldsymbol{\theta}\right)\triangleq D_0\left(\textbf{p}^B,\textbf{q}^B\right) \\ \notag
& = \sum_{i\in\mathcal{I}} \left(\Theta_i+\sum_{m\in\mathcal{M}}\Delta_{i,m}{Q}^{(1)}_{i,m}+\sum_{m\in\mathcal{M}}\Phi_{i,m}{Q}^{(2)}_{i,m}\right)\\  \notag
& \quad \times \ln\frac{ \Theta_i+\sum\limits_{m\in\mathcal{M}}\Delta_{i,m}{Q}^{(1)}_{i,m}+\sum\limits_{m\in\mathcal{M}}\Phi_{i,m}{Q}^{(2)}_{i,m}}{\Psi_i+\sum\limits_{m\in\mathcal{M}}\Gamma_{i,m}{P}^{(1)}_{i,m}+\sum\limits_{m\in\mathcal{M}}\Lambda_{i,m}{P}^{(2)}_{i,m}}\\ \notag
& = \sum_{i\in\mathcal{I}} \left(\Theta_i+\sum_{m\in\mathcal{M}_{\Delta}}\Delta_{i,m}{Q}^{(1)}_{i,m}+\sum_{m\in\mathcal{M}_{\Phi}}\Phi_{i,m}{Q}^{(2)}_{i,m}\right)\\
& \quad \times \ln\frac{ \Theta_i+\sum\limits_{m\in\mathcal{M}_{\Delta}}\Delta_{i,m}{Q}^{(1)}_{i,m}+\sum\limits_{m\in\mathcal{M}_{\Phi}}\Phi_{i,m}{Q}^{(2)}_{i,m}}{\Psi_i+\sum\limits_{m\in\mathcal{M}_{\Gamma}}\Gamma_{i,m}{P}^{(1)}_{i,m}+\sum\limits_{m\in\mathcal{M}_{\Lambda}}\Lambda_{i,m}{P}^{(2)}_{i,m}}.\label{D1_function}
\end{align}
	Note that if $S_m^N(j)=1$ which indicates that the $j$-th IoT device is malicious, then $\gamma_{i,m}^{(j)}p_{i,j,m}^{(1)}=\prod_{l=1}^{L_A-1}p^H_{r^{(j)}_{i,l}}\prod_{l=L_A^+}^{L}p^B_{r^{(j)}_{i,l}}=\Pr\{\hat{\textbf{u}}_j=\textbf{r}_{i}^{(j)}|j\in\mathcal{A},\mathcal{E},\mathcal{H}_1\}$. If  $S_m^N(j)=0$ which indicates that the $j$-th IoT device is honest, then $\gamma_{i,m}^{(j)}p_{i,j,m}^{(1)}=\prod_{l=1}^{L}p^H_{r^{(j)}_{i,l}}=\Pr\{\hat{\textbf{u}}_j=\textbf{r}_{i}^{(j)}|j\not\in\mathcal{A},\mathcal{E},\mathcal{H}_1\}$.
 Hence, we can obtain that
\begin{align}\label{equl_constraint_p1} \notag
&\sum_{i\in\mathcal{I}}\prod_{j=1}^N\gamma_{i,m}^{(j)}p_{i,j,m}^{(1)}=\sum_{i\in\mathcal{I}}\prod_{j=1}^N\Pr\{\hat{\textbf{u}}_j=\textbf{r}_{i}^{(j)}|\mathcal{Y}_j,\mathcal{H}_1\}\\ \notag
&=\sum_{{\textbf{r}}_i^{(1)}\in\mathcal{O}}\sum_{{\textbf{r}}_i^{(2)}\in\mathcal{O}}\cdots\sum_{{\textbf{r}}_i^{(N)}\in\mathcal{O}}\prod_{j=1}^N\Pr\{\hat{\textbf{u}}_j=\textbf{r}_{i}^{(j)}|\mathcal{Y}_j,\mathcal{H}_1\}\\
&=1,
\end{align}where $\mathcal{Y}_j\in\{\{j\in\mathcal{A},\mathcal{E}\},\{j\not\in\mathcal{A},\mathcal{E}\}\}$.
By following a similar argument, we can obtain
\begin{align}\label{equl_constraint_p2} 
\sum_{i\in\mathcal{I}}\prod_{j=1}^N\lambda_{i,m}^{(j)}p_{i,j,m}^{(2)}=1.
\end{align}
By employing (\ref{original_KLD_constraint}), (\ref{equl_constraint_p1}), (\ref{equl_constraint_p2}) and $\sum_{k\in\mathcal{K}}p^H_k=1$ , we can obtain two constraints on ${{P}}_{i,m}^{(1)}$ and ${{P}}_{i,m}^{(2)}$ that 
\begin{align}\label{original_constraint_p1}\notag
\sum_{i\in\mathcal{I}}\Gamma_{i,m}{P}^{(1)}_{i,m}&=P_s(1-\alpha)^{N-n_m}\alpha^{n_m}\sum_{i\in\mathcal{I}}\prod_{j=1}^N\gamma_{i,m}^{(j)}p_{i,j,m}^{(1)}\\
&=P_s(1-\alpha)^{N-n_m}\alpha^{n_m},\forall m\in\mathcal{M}_{\Gamma},
\end{align}
\begin{align}\label{original_constraint_p2} \notag
\sum_{i\in\mathcal{I}}\!\Lambda_{i,m}{P}^{(2)}_{i,m}&=(1-P_s)(1-\alpha)^{N-n_m}\alpha^{n_m}\sum_{i\in\mathcal{I}}\prod_{j=1}^N\lambda_{i,m}^{(j)}p_{i,j,m}^{(2)}\\ 
&=\!(1-P_s)(1-\alpha)^{N-n_m}\alpha^{n_m},\!\forall m\in\mathcal{M}_{\Lambda}.
\end{align}
Similar to (\ref{original_constraint_p1}) and (\ref{original_constraint_p2}), by employing (\ref{original_KLD_constraint}) and $\sum_{k\in\mathcal{O}}q_k^H=1$, we also can obtain two constraints on ${{Q}}_{i,m}^{(1)}$ and ${{Q}}_{i,m}^{(2)}$ that 
\begin{align}\label{original_constraint_q1}
\sum_{i\in\mathcal{I}}\Delta_{i,m}{Q}^{(1)}_{i,m}=P_s(1-\alpha)^{N-n_m}\alpha^{n_m},\forall m\in\mathcal{M}_{\Delta},
\end{align}
\begin{align}\label{original_constraint_q2}
\sum_{i\in\mathcal{I}}\!\Phi_{i,m}{Q}^{(2)}_{i,m}\!\!=\!(1-P_s)(1-\alpha)^{N-n_m}\alpha^{n_m},\!\forall m\in\mathcal{M}_{\Phi}.
\end{align}
Hence, the optimization problem in (\ref{original_KLD}) can be relaxed to the following optimization problem 

\begin{subequations} \label{convex_KLD}
\begin{align} \label{convex_KLD_obj}
\mathop{\min}\limits_{\boldsymbol{\theta}} \quad &D_1\left(\boldsymbol{\theta}\right)\\ 
	\text{s.t.}\quad 	&{P}^{(1)}_{i,m}\in[0,1],\forall i\in\mathcal{I},\forall m\in\mathcal{M}_{\Gamma},\label{convex_KLD_ineq_p1}\\ 
&{P}^{(2)}_{i,m}\in[0,1],\forall i\in\mathcal{I},\forall m\in\mathcal{M}_{\Lambda},\label{convex_KLD_ineq_p2}\\
&{Q}^{(1)}_{i,m}\in[0,1],\forall i\in\mathcal{I},\forall m\in\mathcal{M}_{\Delta},\label{convex_KLD_ineq_q1}\\ 
&{Q}^{(2)}_{i,m}\in[0,1],\forall i\in\mathcal{I},\forall m\in\mathcal{M}_{\Phi},\label{convex_KLD_ineq_q2}\\ \notag
&\text{constraints (\ref{original_constraint_p1})--(\ref{original_constraint_q2}).}
\end{align}
\end{subequations}
Moreover, the optimal value of (\ref{convex_KLD}) provides a lower bound on that of (\ref{original_KLD_obj}), and hence indicates a guaranteed performance for the distributed detection over the BIoT  in the presence of attacks.

\begin{lemma}\label{Lemma_Convex}
	Unlike the original optimization problem in (\ref{original_KLD}) which is nonconvex, the optimization problem in  (\ref{convex_KLD}) is a convex optimization problem.
\end{lemma}
\begin{IEEEproof}
	Refer to Appendix \ref{Proof_Lemma_Convex}.
\end{IEEEproof}
Note that the feasible set of the problem in (\ref{convex_KLD}) is the intersection of the closed sets specified by (\ref{convex_KLD_ineq_p1})--(\ref{convex_KLD_ineq_q2}) and the affine hyperplanes specified by (\ref{original_constraint_p1})--(\ref{original_constraint_q2}). Hence, the feasible set of the problem in (\ref{convex_KLD}) is closed. In addition, since the feasible set of the problem in (\ref{convex_KLD}) is bounded and the objective function in (\ref{convex_KLD}) is continuous, we know from Weirstrass’ Theorem that an optimal solution to the optimization  problem in (\ref{convex_KLD}) exists \cite{arora2004introduction}.  In the following theorem, we provide the analytic form of the optimal solution to the optimization problem in (\ref{convex_KLD}).

\begin{theorem}\label{Theorem_solution}
The optimal solution $\{{P}_{i,m}^{*(1)}\}_{i\in\mathcal{I},m\in\mathcal{M}_{\Gamma}}$, $\{{P}_{i,m}^{*(2)}\}_{i\in\mathcal{I},m\in\mathcal{M}_{\Lambda}}$, $\{{Q}_{i,m}^{*(1)}\}_{i\in\mathcal{I},m\in\mathcal{M}_{\Delta}}$, and $\{{Q}_{i,m}^{*(2)}\}_{i\in\mathcal{I},m\in\mathcal{M}_{\Phi}}$ to the optimization problem in (\ref{convex_KLD}) has the following analytic expressions that $\forall i\in\mathcal{I}$,
\begin{align}\label{Thm_pb1} \notag
&{P}_{i,m}^{*(1)}\!=\\
&\!\min\!\!\left\{\!\!\left[\!\frac{\left({\zeta_{\textbf{p},m}^{*(1)}A^*_i\!-\!\Psi_i\!-\!\!\sum_{m'\in\mathcal{M}_{\Lambda}}\!\!\Lambda_{i,m'}{P}^{*(2)}_{i,m'}}\atop-\sum_{m'\in\mathcal{M}_{\Gamma}\backslash\{m\}}\Gamma_{i,m'}{P}^{*(1)}_{i,m'}\right)}{\Gamma_{i,m}}\!\right]^+\!\!\!,\!1\!\!\right\}\!\!,\!\!\forall m\!\in\!\mathcal{M}_{\Gamma},\\
\label{Thm_pb2} \notag
&{P}_{i,m}^{*(2)}=\\
&\!\min\!\!\left\{\!\!\left[\!\frac{\left({\zeta_{\textbf{p},m}^{*(2)}A^*_i\!-\!\Psi_i\!-\!\!\sum_{m'\in\mathcal{M}_{\Gamma}}\!\!\Gamma_{i,m'}{P}^{*(1)}_{i,m'}\atop-\sum_{m'\in\mathcal{M}_{\Lambda}\backslash\{m\}}\Lambda_{i,m'}{P}^{*(2)}_{i,m'}}\right)}{\Lambda_{i,m}}\!\right]^+\!\!\!,\!1\!\!\right\}\!\!,\!\!\forall m\!\in\!\mathcal{M}_{\Lambda},\\
\label{Thm_qb1}\notag
&{Q}_{i,m}^{*(1)}\!=\\
&\!\min\!\!\left\{\!\!\left[\!\frac{\left(\!{\zeta_{\textbf{q},m}^{*(1)}B^*_{i}\!-\!\Theta_i\!-\!\!\sum_{m'\!\in\!\mathcal{M}_{\Phi}}\!\!\Phi_{i,m'}{Q}^{*(2)}_{i,m'}\atop-\sum_{m'\in\mathcal{M}_{\Delta}\backslash\{m\}}\Delta_{i,m'}{Q}^{*(1)}_{i,m'}}\!\right)}{\Delta_{i,m}}\!\right]^+\!\!\!,\!1\!\!\right\}\!\!,\!\!\forall m\!\in\!\mathcal{M}_{\Delta},
\\
\label{Thm_qb2}\notag
&{Q}_{i,m}^{*(2)}=\\
&\!\min\!\!\left\{\!\!\left[\!\frac{\left(\!{\zeta_{\textbf{q},m}^{*(2)}B^*_{i}\!-\!\Theta_i\!-\!\!\sum_{m'\!\in\!\mathcal{M}_{\Delta}}\!\!\Delta_{i,m'}{Q}^{*(1)}_{i,m'}\atop-\sum_{m'\in\mathcal{M}_{\Phi}\backslash\{m\}}\Phi_{i,m'}{Q}^{*(2)}_{i,m'}}\!\right)}{\Phi_{i,m}}\!\right]^+\!\!\!,\!1\!\!\right\}\!\!,\!\!\forall m\!\in\!\mathcal{M}_{\Phi},
\end{align}
where $[x]^+\triangleq\max\{0,x\}$ for any $x$, $A_i^*\triangleq \Theta_i+\sum_{m\in\mathcal{M}_{\Delta}}\Delta_{i,m}{Q}^{*(1)}_{i,m}+\sum_{m\in\mathcal{M}_{\Phi}}\Phi_{i,m}{Q}^{*(2)}_{i,m}$, and $B_{i}^*\triangleq \Psi_i+\sum_{m\in\mathcal{M}_{\Gamma}}\Gamma_{i,m}{P}^{*(1)}_{i,m}+\sum_{m\in\mathcal{M}_{\Lambda}}\Lambda_{i,m}{P}^{*(2)}_{i,m}$. $\zeta_{\textbf{p},m}^{*(1)}$, $\zeta_{\textbf{p},m}^{*(2)}$, $\zeta_{\textbf{q},m}^{*(1)}$, and $\zeta_{\textbf{q},m}^{*(2)}$ are positive constants which ensure $\sum_{i\in\mathcal{I}}\Gamma_{i,m}{P}_{i,m}^{*(1)}=P_s(1-\alpha)^{N-n_m}\alpha^{n_m}$, $\sum_{i\in\mathcal{I}}\Lambda_{i,m}{P}_{i,m}^{*(2)}=(1-P_s)(1-\alpha)^{N-n_m}\alpha^{n_m}$, $\sum_{i\in\mathcal{I}}\Delta_{i,m}{Q}^{*(1)}_{i,m}=P_s(1-\alpha)^{N-n_m}\alpha^{n_m}$, and $\sum_{i\in\mathcal{I}}\Phi_{i,m}{Q}^{*(2)}_{i,m}=(1-P_s)(1-\alpha)^{N-n_m}\alpha^{n_m}$, respectively.
\end{theorem}
\begin{IEEEproof}
	Refer to Appendix \ref{Proof_Theorem_solution}.
\end{IEEEproof}
Theorem \ref{Theorem_solution} provides the analytic form of the optimal solution to the problem in (\ref{convex_KLD}).  However, as shown in (\ref{Thm_pb1}), (\ref{Thm_pb2}),  (\ref{Thm_qb1}), and (\ref{Thm_qb2}), the analytic expressions of ${P}_{i,m}^{*(1)}$, ${P}_{i,m}^{*(2)}$, ${Q}_{i,m}^{*(1)}$, and ${Q}_{i,m}^{*(2)}$ are coupled with each other. In light of this,  we propose a coordinate descent algorithm to obtain the optimal solution to the problem in (\ref{convex_KLD}) which is summarized in Algorithm \ref{alg_1}.
In order to describe the Algorithm \ref{alg_1} concisely, we first define $\boldsymbol{\theta}^{(t)}$ as a vector stacking $\{{P}_{i,m,(t)}^{(1)}\}_{i\in\mathcal{I},m\in\mathcal{M}_{\Gamma}}$, $\{{{P}}_{i,m,(t)}^{(2)}\}_{ i\in\mathcal{I},m\in\mathcal{M}_{\Lambda}}$, $\{{{Q}}_{i,m,(t)}^{(1)}\}_{ i\in\mathcal{I},m\in\mathcal{M}_{\Delta}}$, and $\{{{Q}}_{i,m,(t)}^{(2)}\}_{i\in\mathcal{I}, m\in\mathcal{M}_{\Phi}}$, and define $\textbf{P}_{m,(t)}^{(1)}$, $\forall m\in\mathcal{M}_{\Gamma}$, $\textbf{P}_{m,(t)}^{(2)}$, $\forall m\in\mathcal{M}_{\Lambda}$, $\textbf{Q}_{m,(t)}^{(1)}$, $\forall m\in\mathcal{M}_{\Delta}$, and $\textbf{Q}_{m,(t)}^{(2)}$, $\forall m\in\mathcal{M}_{\Phi}$ as vectors stacking $\{{P}_{i,m,(t)}^{(1)}\}_{i\in\mathcal{I}}$, $\{{P}_{i,m,(t)}^{(2)}\}_{i\in\mathcal{I}}$, $\{{Q}_{i,m,(t)}^{(1)}\}_{i\in\mathcal{I}}$, and
$\{{Q}_{i,m,(t)}^{(2)}\}_{i\in\mathcal{I}}$, respectively. We also define
\begin{align}\label{Alg_A}A_{i}\triangleq \Theta_i+\sum\limits_{m\in\mathcal{M}_{\Delta}}\Delta_{i,m}{Q}^{(1)}_{i,m}+\sum\limits_{m\in\mathcal{M}_{\Phi}}\Phi_{i,m}{Q}^{(2)}_{i,m},\end{align}
\begin{align}\label{Alg_B}B_{i}\triangleq \Psi_i+\sum\limits_{m\in\mathcal{M}_{\Gamma}}\Gamma_{i,m}{P}^{(1)}_{i,m}+\sum\limits_{m\in\mathcal{M}_{\Lambda}}\Lambda_{i,m}{P}^{(2)}_{i,m}.\end{align}

\begin{figure}[H]\centering 
	\includegraphics[width=2.5in]{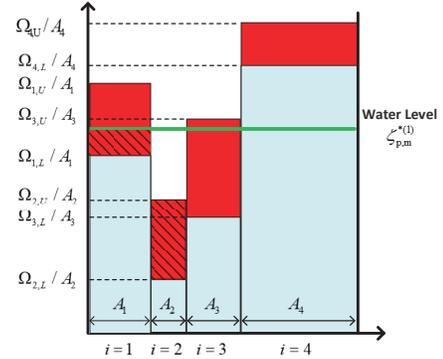}    
	\caption{The diagram of the capped water-filling method.}
	\label{fig_water} 
\end{figure}

\begin{theorem} \label{Theorem_CoordinateDescent}
Algorithm \ref{alg_1} can converge to the globally optimal solution to the problem in (\ref{convex_KLD}).
\end{theorem}
\begin{IEEEproof}
	Since the problem in (\ref{convex_KLD}) is a convex optimization problem and the objective function (\ref{convex_KLD_obj}) is twice continuously differentiable, the coordinate descent method can converge to its optimal solution \cite{luo1992convergence}.\cbk
\end{IEEEproof}
The constants ${\zeta_{\textbf{p},m}^{*(1)}}$, ${\zeta_{\textbf{p},m}^{*(2)}}$, ${\zeta_{\textbf{q},m}^{*(1)}}$, and ${\zeta_{\textbf{q},m}^{*(2)}}$ in  (\ref{Thm_pb1})--(\ref{Thm_qb2}) can be obtained by employing a variant of the water-filling procedure which is referred to as a capped water-filling method. We take Step \ref{alg_update_p1} of Algorithm \ref{alg_1} as an example to describe this procedure, which is illustrated in Fig. \ref{fig_water}. Note that in Step \ref{alg_update_p1} of Algorithm \ref{alg_1}, $\{\textbf{P}_{m'}^{(1)}\}_{m'\in\mathcal{M}_{\Gamma}\backslash\{m\}},\{\textbf{P}_{m'}^{(2)}\}_{m'\in\mathcal{M}_{\Lambda}},\{\textbf{Q}_{m'}^{(1)}\}_{m'\in\mathcal{M}_{\Delta}},\{\textbf{Q}_{m'}^{(2)}\}_{m'\in\mathcal{M}_{\Phi}}$ are considered as known constants. For each $i\in\mathcal{I}$, we first calculate $A_i$ by using (\ref{Alg_A}). Then we calculate the minimum $B_i$ by using (\ref{Alg_B}) with $P_{i,m}^{(1)}=0$ and the maximum $B_i$ by using (\ref{Alg_B}) with $P_{i,m}^{(1)}=1$, which are denoted as $\Omega_{i,L}$ and $\Omega_{i,U}$, respectively. In Fig. \ref{fig_water} where $|\mathcal{I}|=4$, for each $i\in\mathcal{I}$, we first draw a rectangle with the area, the base, and the height equal to $\Omega_{i,U}$, $A_i$, and ${\Omega_{i,U}}/{A_i}$, respectively.

\begin{algorithm}
	\caption{}
	\label{alg_1}
	\begin{algorithmic}[1]

		\State{\textbf{Initialization:}}\label{alg_init} Arbitrarily initialize the  PMFs \cbk $\textbf{p}^B$ and $\textbf{q}^B$ in (\ref{equ_qb}) and then initialize ${\textbf{P}}_{m}^{(1)}$, $\forall m\in\mathcal{M}_{\Gamma}$, ${\textbf{P}}_{m}^{(2)}$, $\forall m\in\mathcal{M}_{\Lambda}$, ${\textbf{Q}}_{m}^{(1)}$, $\forall m\in\mathcal{M}_{\Delta}$, and ${\textbf{Q}}_{m}^{(2)}$, $\forall m\in\mathcal{M}_{\Phi}$ in $\boldsymbol{\theta}$ by using (\ref{def_P1}), (\ref{def_P2}), (\ref{def_Q1}), and (\ref{def_Q2}), respectively. Update $\boldsymbol{\theta}^{(0)}$ with $\boldsymbol{\theta}$. Set $t=0$, and choose a small constant $\epsilon>0$.
		\State{Calculate $D_{(0)}\triangleq D_1(\boldsymbol{\theta}^{(0)})$ by using (\ref{D1_function})}.
		\State{\textbf{Repeat}}
		\For {$ m=1,...,2^N-1$}
		\State Calculate $n_m$ by using (\ref{equ_nm}). 
		\For {$ i=1,...,|\mathcal{R}|$} \Comment{minimize over ${\textbf{P}}_{m}^{(1)}$}

\If{$\Gamma_{i,m}\not=0$}
\State Calculate $A_i$ by using (\ref{Alg_A}) with $\boldsymbol{\theta}\leftarrow\boldsymbol{\theta}^{(t)}$. 
\State  {Update $P_{i,m,(t+1)}^{(1)}$ by using right-hand side of }\label{alg_update_p1}
\Statex \qquad \qquad \; (\ref{Thm_pb1}) with $A^*_{i}\!\leftarrow\! A_{i}$,
     ${P}^{*(2)}_{i,m'}\leftarrow{P}^{(2)}_{i,m'},m'\in\mathcal{M}_{\Lambda}$, 
\Statex \qquad \qquad \; and ${P}^{*(1)}_{i,m'}\leftarrow{P}^{(1)}_{i,m'},m'\in\mathcal{M}_{\Gamma}\backslash\{m\}$.\cbk
\EndIf
		\EndFor 
\State ${\textbf{P}}_{m}^{(1)}\leftarrow{\textbf{P}}_{m,(t+1)}^{(1)}$ and update $\boldsymbol{\theta}^{(t)}$ by replacing ${\textbf{P}}_{m,(t)}^{(1)}$ 
\Statex \quad\; with ${\textbf{P}}_{m,(t+1)}^{(1)}$.
		\For {$ i=1,...,|\mathcal{R}|$} \Comment{minimize over ${\textbf{P}}_{m}^{(2)}$}
		
\If{$\Lambda_{i,m}\not=0$}
\State Calculate $A_i$ by using (\ref{Alg_A}) with $\boldsymbol{\theta}\leftarrow\boldsymbol{\theta}^{(t)}$.
\State\label{alg_update_p2} Update $P_{i,m,(t+1)}^{(2)}$ by using right-hand side of 
\Statex \qquad \qquad \; (\ref{Thm_pb2}) with $A^*_{i}\!\leftarrow \!A_{i}$, ${P}^{*(1)}_{i,m'}\leftarrow{P}^{(1)}_{i,m'},m'\in\mathcal{M}_{\Gamma}$,
\Statex \qquad \qquad \;      and ${P}^{*(2)}_{i,m'}\leftarrow{P}^{(2)}_{i,m'},m'\in\mathcal{M}_{\Lambda}\backslash\{m\}$.\cbk
\EndIf
		\EndFor
\State ${\textbf{P}}_{m}^{(2)}\leftarrow{\textbf{P}}_{m,(t+1)}^{(2)}$ and update $\boldsymbol{\theta}^{(t)}$ by replacing ${\textbf{P}}_{m,(t)}^{(2)}$
\Statex \quad\; with ${\textbf{P}}_{m,(t+1)}^{(2)}$.
		\For {$ i=1,...,|\mathcal{R}|$} \Comment{minimize over ${\textbf{Q}}_{m}^{(1)}$}
		\If{$\Delta_{i,m}\not=0$}
\State Calculate $B_i$ by using (\ref{Alg_B}) with $\boldsymbol{\theta}\leftarrow\boldsymbol{\theta}^{(t)}$.
\State \label{alg_update_q1}Update $Q_{i,m,(t+1)}^{(1)}$ by using right-hand side of \Statex \qquad \qquad \; (\ref{Thm_qb1}) with $B^*_{i}\!\leftarrow\! B_{i}$, ${Q}^{*(2)}_{i,m'}\!\leftarrow\!{Q}^{(2)}_{i,m'},m'\in\mathcal{M}_{\Phi}$,
\Statex \qquad \qquad \; and ${Q}^{*(1)}_{i,m'}\!\leftarrow\!{Q}^{(1)}_{i,m'},m'\in\mathcal{M}_{\Delta}\backslash\{m\}$.\cbk
\EndIf
\EndFor
\State ${\textbf{Q}}_{m}^{(1)}\leftarrow{\textbf{Q}}_{m,(t+1)}^{(1)}$ and update $\boldsymbol{\theta}^{(t)}$ by replacing ${\textbf{Q}}_{m,(t)}^{(1)}$
\Statex \quad\; with ${\textbf{Q}}_{m,(t+1)}^{(1)}$.
		\For {$ i=1,...,|\mathcal{R}|$} \Comment{minimize over ${\textbf{Q}}_{m}^{(2)}$}
\If{$\Phi_{i,m}\not=0$}
		\State Calculate $B_i$ by using (\ref{Alg_B}) with $\boldsymbol{\theta}=\boldsymbol{\theta}^{(t)}$. 
\State\label{alg_update_q2} Update $Q_{i,m,(t+1)}^{(2)}$ by using right-hand side of \Statex \qquad \qquad \;(\ref{Thm_qb2}) with $B^*_{i}\!\leftarrow\! B_{i}$, ${Q}^{*(1)}_{i,m'}\!\leftarrow\!{Q}^{(1)}_{i,m'},m'\in\mathcal{M}_{\Delta}$,
\Statex \qquad \qquad \;  and ${Q}^{*(2)}_{i,m'}\leftarrow{Q}^{(2)}_{i,m'},m'\in\mathcal{M}_{\Phi}\backslash\{m\}$.\cbk
\EndIf
	\EndFor
\State ${\textbf{Q}}_{m}^{(2)}\leftarrow{\textbf{Q}}_{m,(t+1)}^{(2)}$ and update $\boldsymbol{\theta}^{(t)}$ by replacing ${\textbf{Q}}_{m,(t)}^{(2)}$ 
\Statex \quad\; with ${\textbf{Q}}_{m,(t+1)}^{(2)}$.
\EndFor
\State $\boldsymbol{\theta}^{(t+1)}\leftarrow\boldsymbol{\theta}^{(t)}.$\label{alg_update_t}
\State $t\leftarrow t+1$. Calculate $D_{(t)}\triangleq D_1(\boldsymbol{\theta}^{(t)})$ by using (\ref{D1_function}).
		\State {\textbf{Until} $|D_{(t)}-D_{(t-1)}|<\epsilon$}.
		\State{\textbf{Output:}} $D_1(\boldsymbol{\theta}^{(t)})$, $\boldsymbol{\theta}^{(t)}$.   
	\end{algorithmic}  
\end{algorithm}
Then we start to gradually increase the level $\zeta_{\textbf{p},m}^{*(1)}$ of ``water''  from its initial value zero. Once $\zeta_{\textbf{p},m}^{*(1)}$ reaches $\Omega_{i,L}/A_i$, the water begins to fill the portion of the $i$-th rectangle that is above $\Omega_{i,L}/A_i$ (see the red parts of the rectangles in Fig. \ref{fig_water}). As $\zeta_{\textbf{p},m}^{*(1)}$ increases, if $\zeta_{\textbf{p},m}^{*(1)}$ passes $\Omega_{i,U}/A_i$, then the portion of the $i$-th rectangle that is above  $\Omega_{i,L}/A_i$ has been filled up, and hence the water can no longer fill the $i$-th rectangle. The water in the rectangles is indicated by the rectangles filled with inclined lines in Fig. \ref{fig_water}. The process of increasing $\zeta_{\textbf{p},m}^{*(1)}$ is stopped when the total area of the rectangles filled with inclined lines is equal to $P_s(1-\alpha)^{N-n_m}\alpha^{n_m}$,  and the corresponding value of $\zeta_{\textbf{p},m}^{*(1)}$ is the desired value of $\zeta_{\textbf{p},m}^{*(1)}$ in (\ref{Thm_pb1}) which is depicted by the green line in Fig. \ref{fig_water}.
The procedures to solve (\ref{Thm_pb2}), (\ref{Thm_qb1}), and (\ref{Thm_qb2}) are similar to this capped water-filling procedure.

\section{Numerical Results}\label{Section_simulation}

The optimal objective of the problem in (\ref{convex_KLD}) yielded by Algorithm \ref{alg_1} provides a guarantee on the detection performance of the BIoT in the presence of attacks. To corroborate this performance guarantee, we numerically investigate the optimal solution to the problem in (\ref{original_KLD}) by using the projected gradient descent (PGD) method with multiple initial points \cite{daubechies2008accelerated}. In all the following simulation results, we consider the scenario where $\mathcal{K}=\{0,1\}$, $N=4$, $\textbf{p}^H=[0.1,0.9]$, $\textbf{q}^H=[0.9, 0.1]$, and $L=4$.

\begin{figure}[H]\centering   
	\includegraphics[width=2.5in]{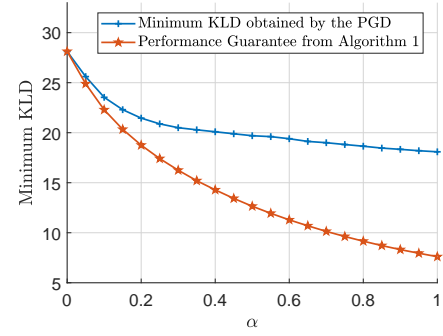}    
	\caption{Comparison between the minimum KLD and its guarantee.}
	\label{fig_KLD1}
\end{figure}
As the probability $\alpha$ that an IoT device has been hacked into varies from $0$ to $1$, Fig. \ref{fig_KLD1} depicts the minimum KLD obtained by using the PGD with multiple initial points and the performance guarantee yielded by Algorithm \ref{alg_1} for the case where  $L_0=3,L_A=3$\cbk, and $P_s= 0.0118$, which are marked with `+'s and pentagrams, respectively.
It is seen from Fig. \ref{fig_KLD1} that the performance guarantee yielded by Algorithm \ref{alg_1} provides a lower bound on the minimum KLD obtained by the PGD, which  provides valuable insights into the worst detection performance of the BIoT in adversarial environments.

\begin{figure}[H]\centering   
	\includegraphics[width=2.5in]{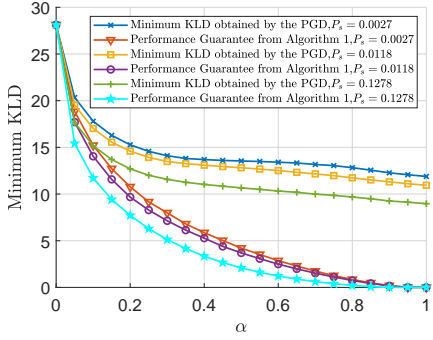}    
	\caption{Comparison between the minimum KLD and its guarantee for different $P_s$.}
	\label{fig_KLD2}
\end{figure}

Next, we investigate how the value of $P_s$ impacts the minimum KLD and the performance guarantee obtained from Algorithm \ref{alg_1}. As $\alpha$ varies from $0$ to $1$, Fig. \ref{fig_KLD2} depicts the minimum KLD obtained by using the PGD with multiple initial points and the performance guarantee yielded by Algorithm \ref{alg_1} with different $P_s$ for the case where $L_0=2$ and $L_A=2$\cbk. The blue curve marked with `x's and the orange curve marked with triangles illustrate the minimum KLD obtained by using the PGD and the performance guarantee obtained from Algorithm \ref{alg_1} when $P_s=0.0027$. The yellow curve marked with squares and the purple curve marked with circles illustrate the minimum KLD obtained by using the PGD and the performance guarantee obtained from Algorithm \ref{alg_1} when $P_s=0.0118$.  The green curve marked with `+'s and the cyan curve marked with pentagrams illustrate the minimum KLD obtained by using the PGD and the performance guarantee obtained from Algorithm \ref{alg_1} when $P_s=0.1278$. \cbk
 It is seen from Fig. \ref{fig_KLD2} that the performance guarantees yielded by Algorithm \ref{alg_1} still provide lower bounds on the minimum KLDs obtained by the PGD for different $P_s$. In addition, Fig. \ref{fig_KLD2} shows that as $P_s$ increases, the minimum KLD and the performance guarantee both decrease. This is because as $P_s$ increases, the probability of successfully launching a DSA becomes greater, and hence, the detection performance of the BIoT is degraded to a larger extent.
\begin{figure}[H]\centering   
	\includegraphics[width=2.5in]{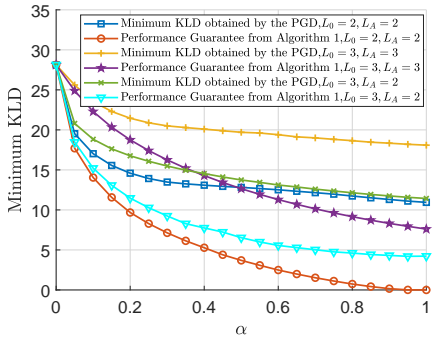}    
	\caption{Comparison between the minimum KLD and its guarantee for different $L_0$ and $L_A$.}
	\label{fig_KLD3}
\end{figure}

Fig. \ref{fig_KLD3} depicts the minimum KLD obtained by using the PGD with multiple initial points and the performance guarantee yielded by Algorithm \ref{alg_1} with different $L_0$ and $L_A$ for the case where $P_s= 0.0118$. The blue curve marked with `+'s and the orange curve marked with triangles illustrate the minimum KLD obtained by using the PGD and the performance guarantee obtained from Algorithm \ref{alg_1} when $L_0=2$ and $L_A=2$. The yellow curve marked with squares and the purple curve marked with pentagrams illustrate the minimum KLD obtained by using the PGD and the performance guarantee obtained from Algorithm \ref{alg_1} when $L_0=3$ and $L_A=3$.  The green curve marked with `x's and the cyan curve marked with triangles illustrate the minimum KLD obtained by using the PGD and the performance guarantee obtained from Algorithm \ref{alg_1} when $L_0=3$ and $L_A=2$. \cbk
Fig. \ref{fig_KLD3} also shows that the performance guarantee yielded by Algorithm \ref{alg_1} provides valid lower bounds on the minimum KLD obtained by the PGD. In addition, Fig. \ref{fig_KLD3} shows that as $L_0$ and $L_A$ increase, the minimum KLD and the performance guarantee both increase. This is because as $L_0$ and $L_A$ increase, the numbers of counterfeit blocks in the counterfeit branch and the authentic branch both decrease. Therefore, the detection performance of the BIoT becomes better.

\section{Conclusion}\label{Section_conclusion}
{ In this paper, we considered distributed detection over a BIoT in the presence of attacks which jointly exploit the vulnerability of IoT devices and that of the blockchain employed in the BIoT. The pursuit of a detection performance guarantee for the BIoT has been cast as a relaxed convex optimization problem, and the analytic expression for the solution to the relaxed convex optimization problem has been derived. Moreover, based on a capped water-filling method, we have developed a coordinate descent algorithm with guaranteed convergence to solve the relaxed convex optimization problem. }

\appendices
\section{Proof of Lemma \ref{Lemma_Convex}} \label{Proof_Lemma_Convex}
Let $\tilde{\boldsymbol{\theta}}$ denote a vector stacking $\{\tilde{{P}}_{i,m}^{(1)}\}_{i\in\mathcal{I},m\in\mathcal{M}_{\Gamma}},\{\tilde{{P}}_{i,m}^{(2)}\}_{i\in\mathcal{I},m\in\mathcal{M}_{\Lambda}},\{\tilde{{Q}}_{i,m}^{(1)}\}_{i\in\mathcal{I},m\in\mathcal{M}_{\Delta}}$, and $\{\tilde{{Q}}_{i,m}^{(2)}\}_{i\in\mathcal{I},m\in\mathcal{M}_{\Phi}}$. Similarly, we define $\check{\boldsymbol{\theta}}$ as a vector stacking $\{\check{{P}}_{i,m}^{(1)}\}_{i\in\mathcal{I},m\in\mathcal{M}_{\Gamma}},\{\check{{P}}_{i,m}^{(2)}\}_{i\in\mathcal{I},m\in\mathcal{M}_{\Lambda}},\{\check{{Q}}_{i,m}^{(1)}\}_{i\in\mathcal{I},m\in\mathcal{M}_{\Delta}}$, and $\{\check{{Q}}_{i,m}^{(2)}\}_{i\in\mathcal{I},m\in\mathcal{M}_{\Phi}}$, and $\boldsymbol{\theta} \triangleq \mu\tilde{\boldsymbol{\theta}}+(1-\mu)\check{\boldsymbol{\theta}}$ for some $\mu\in[0,1]$. From (\ref{D1_function}), we can obtain 
\begin{align} \notag
&D_1\left(\boldsymbol{\theta}\right) \\ \notag
& =\sum_{i\in\mathcal{I}}\left\{\Theta_i+\sum\limits_{m\in\mathcal{M}_{\Delta}}\Delta_{i,m}\left[\mu\tilde{Q}^{(1)}_{i,m}+(1-\mu)\check{Q}^{(1)}_{i,m}\right]\right.\\ \notag
&\qquad\qquad+\left.\sum\limits_{m\in\mathcal{M}_{\Phi}}\Phi_{i,m}\left[\mu\tilde{Q}^{(2)}_{i,m}+(1-\mu)\check{Q}^{(2)}_{i,m}\right]\right\}\\ \notag
&\quad\times\ln\frac{\left\{{\Theta_i+\sum\limits_{m\in\mathcal{M}_{\Delta}}\Delta_{i,m}\left[\mu\tilde{Q}^{(1)}_{i,m}+(1-\mu)\check{Q}^{(1)}_{i,m}\right]\atop+\sum\limits_{m\in\mathcal{M}_{\Phi}}\Phi_{i,m}\left[\mu\tilde{Q}^{(2)}_{i,m}+(1-\mu)\check{Q}^{(2)}_{i,m}\right]}\right\}}{\left\{{\Psi_i+\sum\limits_{m\in\mathcal{M}_{\Gamma}}\Gamma_{i,m}\left[\mu\tilde{P}^{(1)}_{i,m}+(1-\mu)\check{P}^{(1)}_{i,m}\right]\atop+\sum\limits_{m\in\mathcal{M}_{\Lambda}}\Lambda_{i,m}\left[\mu\tilde{P}^{(2)}_{i,m}+(1-\mu)\check{P}^{(2)}_{i,m}\right]}\right\}}\\ \notag
& \le \sum_{i\in\mathcal{I}}\mu\left(\Theta_i+\sum\limits_{m\in\mathcal{M}_{\Delta}}\Delta_{i,m}\tilde{Q}^{(1)}_{i,m}+\sum\limits_{m\in\mathcal{M}_{\Phi}}\Phi_{i,m}\check{Q}^{(2)}_{i,m}\right) \\ \notag 
&\times \ln\frac{\mu\left(\Theta_i+\sum\limits_{m\in\mathcal{M}_{\Delta}}\Delta_{i,m}\tilde{Q}^{(1)}_{i,m}+\sum\limits_{m\in\mathcal{M}_{\Phi}}\Phi_{i,m}\tilde{Q}^{(2)}_{i,m}\right)}{\mu\left(\Psi_i+\sum\limits_{m\in\mathcal{M}_{\Gamma}}\Gamma_{i,m}\tilde{P}^{(1)}_{i,m}+\sum\limits_{m\in\mathcal{M}_{\Lambda}}\Lambda_{i,m}\tilde{P}^{(2)}_{i,m}\right)} \\\notag
&+\sum_{i\in\mathcal{I}}(1-\mu)\!\!\left(\!\Theta_i\!+\!\!\sum\limits_{m\in\mathcal{M}_{\Delta}}\!\Delta_{i,m}\check{Q}^{(1)}_{i,m}\!+\!\!\sum\limits_{m\in\mathcal{M}_{\Phi}}\!\!\Phi_{i,m}\check{Q}^{(2)}_{i,m}\right)\\
  \label{Lemma_Convex_Proof_1}
&\times\!\ln\!\frac{(1-\mu)\!\!\left(\!{\Theta_i\!+\!\!\sum\limits_{m\in\mathcal{M}_{\Delta}}\!\!\Delta_{i,m}\check{Q}^{(1)}_{i,m}\! +\!\!\sum\limits_{m\in\mathcal{M}_{\Phi}}\!\!\Phi_{i,m}\check{Q}^{(2)}_{i,m}}\!\right)}{(1-\mu)\!\!\left(\Psi_i\!+\!\!\sum\limits_{m\in\mathcal{M}_{\Gamma}}\!\!\Gamma_{i,m}\check{P}^{(1)}_{i,m}\!+\!\!\sum\limits_{m\in\mathcal{M}_{\Lambda}}\!\!\Lambda_{i,m}\check{P}^{(2)}_{i,m}\!\right)}\\
& = \mu D_1\Big(\tilde{\boldsymbol{\theta}}\Big)+(1-\mu)D_1\Big(\check{\boldsymbol{\theta}}\Big),\label{Lemma_Convex_Proof_2}
\end{align}
where the inequality in (\ref{Lemma_Convex_Proof_1}) comes from the fact that for any nonnegative $a_1,a_2,...,a_n$ and $b_1,b_2,...,b_n$, we have $\sum_{i=1}^n a_i\log\frac{a_i}{b_i}\ge (\sum_{i=1}^n a_i)\log\frac{\sum_{i=1}^n a_i}{\sum_{i=1}^n b_i}$. Therefore, 
$D_1(\boldsymbol{\theta})$ is a convex function with respect to $\boldsymbol{\theta}$. Moreover, since the inequality constraint functions in (\ref{convex_KLD_ineq_p1})--(\ref{convex_KLD_ineq_q2}) are convex and the equality constraint functions in (\ref{original_constraint_p1})--(\ref{original_constraint_q2}) are affine functions of $\boldsymbol{\theta}$, the optimization problem in (\ref{convex_KLD}) is a convex optimization problem.

\section{Proof of Theorem \ref{Theorem_solution}} \label{Proof_Theorem_solution}
Define $\textbf{P}_m^{*(1)}$, $\forall m\in\mathcal{M}_{\Gamma}$, $\textbf{P}_m^{*(2)}$, $\forall m\in\mathcal{M}_{\Lambda}$, $\textbf{Q}_m^{*(1)}$, $\forall m\in\mathcal{M}_{\Delta}$, and $\textbf{Q}_m^{*(2)}$, $\forall m\in\mathcal{M}_{\Phi}$  as vectors stacking $\{{P}_{i,m}^{*(1)}\}_{i\in\mathcal{I}}$, $\{{P}_{i,m}^{*(2)}\}_{i\in\mathcal{I}}$, $\{{Q}_{i,m}^{*(1)}\}_{i\in\mathcal{I}}$, and $\{{Q}_{i,m}^{*(2)}\}_{i\in\mathcal{I}}$, respectively.
Let $D^*_1({\textbf{Q}}_{m}^{(1)})\triangleq D_1(\boldsymbol{\theta})|{\{\textbf{Q}_{m'}^{(1)}=\textbf{Q}_{m'}^{*(1)}\}_{m'\in\mathcal{M}_{\Delta}\backslash\{m\}},\{\textbf{P}_{m'}^{(1)}=\textbf{P}_{m'}^{*(1)}\}_{m'\in\mathcal{M}_{\Gamma}},}$ ${\{\textbf{P}_{m'}^{(2)}=\textbf{P}_{m'}^{*(2)}\}_{m'\in\mathcal{M}_{\Lambda}},\{\textbf{Q}_{m'}^{(2)}=\textbf{Q}_{m'}^{*(2)}\}_{m'\in\mathcal{M}_{\Phi}}}$ which indicates that we only consider ${\textbf{Q}}_{m}^{(1)}$ in $\boldsymbol{\theta}$ as the variable of $D^*_1({\textbf{Q}}_{m}^{(1)})$, and the other parameters in $\boldsymbol{\theta}$ are set to be the globally optimal solutions to the optimization problem in (\ref{convex_KLD}). Similarly, we define 
$D^*_2({\textbf{Q}}_{m}^{(2)})\triangleq D_1(\boldsymbol{\theta})|{\{\textbf{Q}_{m'}^{(2)}=\textbf{Q}_{m'}^{*(2)}\}_{m'\in\mathcal{M}_{\Phi}\backslash\{m\}},\{\textbf{P}_{m'}^{(1)}=\textbf{P}_{m'}^{*(1)}\}_{m'\in\mathcal{M}_{\Gamma}},}$ ${\{\textbf{P}_{m'}^{(2)}=\textbf{P}_{m'}^{*(2)}\}_{m'\in\mathcal{M}_{\Lambda}},\{\textbf{Q}_{m'}^{(1)}=\textbf{Q}_{m'}^{*(1)}\}_{m'\in\mathcal{M}_{\Delta}}}$, 
$D^*_3({\textbf{P}}_{m}^{(1)})\triangleq D_1(\boldsymbol{\theta})|{\{\textbf{P}_{m'}^{(1)}=\textbf{P}_{m'}^{*(1)}\}_{m'\in\mathcal{M}_{\Gamma}\backslash\{m\}},\{\textbf{P}_{m'}^{(2)}=\textbf{P}_{m'}^{*(2)}\}_{m'\in\mathcal{M}_{\Lambda}},}$ ${\{\textbf{Q}_{m'}^{(1)}=\textbf{Q}_{m'}^{*(1)}\}_{m'\in\mathcal{M}_{\Delta}},\{\textbf{Q}_{m'}^{(2)}=\textbf{Q}_{m'}^{*(2)}\}_{m'\in\mathcal{M}_{\Phi}}}$, and $D^*_4({\textbf{P}}_{m}^{(2)})\triangleq D_1(\boldsymbol{\theta})|{\{\textbf{P}_{m'}^{(2)}=\textbf{P}_{m'}^{*(2)}\}_{m'\in\mathcal{M}_{\Lambda}\backslash\{m\}},\{\textbf{P}_{m'}^{(1)}=\textbf{P}_{m'}^{*(1)}\}}$ $_{m'\in\mathcal{M}_{\Gamma}},{\{\textbf{Q}_{m'}^{(1)}=\textbf{Q}_{m'}^{*(1)}\}_{m'\in\mathcal{M}_{\Delta}},\{\textbf{Q}_{m'}^{(2)}=\textbf{Q}_{m'}^{*(2)}\}_{m'\in\mathcal{M}_{\Phi}}}$.

Since $\{{P}_{i,m}^{*(1)}\}_{i\in\mathcal{I},m\in\mathcal{M}_{\Gamma}}$, $\{{P}_{i,m}^{*(2)}\}_{i\in\mathcal{I},m\in\mathcal{M}_{\Lambda}}$, $\{{Q}_{i,m}^{*(1)}\}_{i\in\mathcal{I},m\in\mathcal{M}_{\Delta}}$, and $\{{Q}_{i,m}^{*(2)}\}_{i\in\mathcal{I},m\in\mathcal{M}_{\Phi}}$ are the globally optimal solutions  to the optimization problem in (\ref{convex_KLD}), we know that
\begin{align}\label{Appendix_q1*}
{\textbf{Q}}_{m}^{*(1)}&=\mathop{\text{arg\,min}}\limits_{{\textbf{Q}}_{m}^{(1)}} D^*_1\left({\textbf{Q}}_{m}^{(1)}\right),\forall m\in\mathcal{M}_{\Delta}\\\notag
&\text{s.t.}\quad \text{constraints (\ref{original_constraint_q1}), (\ref{convex_KLD_ineq_q1}).}\\
\label{Appendix_q2*}
{\textbf{Q}}_{m}^{*(2)}&=\mathop{\text{arg\,min}}\limits_{{\textbf{Q}}_{m}^{(2)}} D^*_2\left({\textbf{Q}}_{m}^{(2)}\right),\forall m\in\mathcal{M}_{\Phi}\\\notag
&\text{s.t.}\quad \text{constraints (\ref{original_constraint_q2}), (\ref{convex_KLD_ineq_q2}).}\\
\label{Appendix_p1*}
{\textbf{P}}_{m}^{*(1)}&=\mathop{\text{arg\,min}}\limits_{{\textbf{P}}_{m}^{(1)}} D^*_3\left({\textbf{P}}_{m}^{(1)}\right),\forall m\in\mathcal{M}_{\Gamma}\\\notag
&\text{s.t.}\quad \text{constraints (\ref{original_constraint_p1}), (\ref{convex_KLD_ineq_p1}).}\\
\label{Appendix_p2*}
{\textbf{P}}_{m}^{*(2)}&=\mathop{\text{arg\,min}}\limits_{{\textbf{P}}_{m}^{(2)}} D^*_4\left({\textbf{P}}_{m}^{(2)}\right),\forall m\in\mathcal{M}_{\Lambda}\\\notag
&\text{s.t.}\quad \text{constraints (\ref{original_constraint_p2}), (\ref{convex_KLD_ineq_p2}).}\end{align}
From (\ref{D1_function}), we can obtain the following property of the optimal solution.
\begin{claim}\label{claim_property}The globally optimal solution to (\ref{convex_KLD}) cannot give rise to that $A^*_i>0$ and $B^*_i=0$ for any $i\in\mathcal{I}$.
\end{claim}
\begin{IEEEproof}[Proof of Claim \ref{claim_property}]
We prove this claim by contradiction. Suppose that the globally optimal solution to (\ref{convex_KLD}), denoted by $\boldsymbol{\theta}^*$, gives rise to that $A^*_{i'}>0$ and $B^*_{i'}=0$ for some $i'\in\mathcal{I}$. We know from (\ref{D1_function}) and the definitions of $A^*_{i}$ and $B^*_{i}$ in Theorem \ref{Theorem_solution} that $D_1(\boldsymbol{\theta}^*)=\infty$. From (\ref{DM_under_H1_expand}), (\ref{def_Gamma}), (\ref{def_Lambda}), (\ref{equ_gamma}), (\ref{equ_lambda}), and the definition of $B^*_{i}$, we know that $B^*_{i'}=0$ only if $\alpha=1$ and $\Gamma_{i',2^N-1}P_{i',2^N-1}^{*(1)}=\Lambda_{i',2^N-1}P_{i',2^N-1}^{*(2)}=0$. Note that if $\alpha=1$ and $P_s>0$, then $\Gamma_{i',2^N-1}>0$, and hence $P_{i',2^N-1}^{*(1)}=0$. If $\alpha=1$ and $P_s=0$, then $\Lambda_{i',2^N-1}>0$, and hence $P_{i',2^N-1}^{*(2)}=0$. 
We first arbitrarily pick an $i^{\dagger}\in\{i\in\mathcal{I}|P_{i,2^N-1}^{*(1)}>0\}$ if $P_s>0$ (or $i^{\dagger}\in\{i\in\mathcal{I}|P_{i,2^N-1}^{*(2)}>0\}$ if $P_s=0$). 
It's worth mentioning that when $\alpha=1$ we can always find such an $i^{\dagger}$ due to the constraint in (\ref{original_constraint_p1}) that $\sum_{i\in\mathcal{I}}\Gamma_{i,2^N-1}{P}^{*(1)}_{i,2^N-1}=P_s>0$ if $P_s>0$ (or (\ref{original_constraint_p2}) that $\sum_{i\in\mathcal{I}}\Lambda_{i,2^N-1}{P}^{*(2)}_{i,2^N-1}=1-P_s>0$ if $P_s=0$). We increase the value of $P_{i',2^N-1}^{*(1)}$ by some constant $\epsilon$ where $0<\epsilon<\min\{1,(\Gamma_{i^{\dagger},2^N-1}P_{i^{\dagger},2^N-1}^{*(1)})/\Gamma_{i',2^N-1}\}$ if $P_s>0$ (or increase the value of $P_{i',2^N-1}^{*(2)}$ by some constant $\epsilon$ where $0<\epsilon<\min\{1,(\Lambda_{i^{\dagger},2^N-1}P_{i^{\dagger},2^N-1}^{*(2)})/\Lambda_{i',2^N-1}\}$ if $P_s=0$), and decrease the value of $P_{i^{\dagger},2^N-1}^{*(1)}$ by the constant $\epsilon\Gamma_{i',2^N-1}/\Gamma_{i^{\dagger},2^N-1}$ if $P_s>0$ (or decrease the value of $P_{i^{\dagger},2^N-1}^{*(2)}$ by the constant $\epsilon\Lambda_{i',2^N-1}/\Lambda_{i^{\dagger},2^N-1}$ if $P_s=0$) so that the value of $\sum_{i\in\mathcal{I}}\Gamma_{i,2^N-1}P_{i,2^N-1}^{*(1)}$ (or $\sum_{i\in\mathcal{I}}\Lambda_{i,2^N-1}P_{i,2^N-1}^{*(2)}$) doesn't change and hence the constraint in (\ref{original_constraint_p1}) (or (\ref{original_constraint_p2})) still holds. By doing so, the adjusted $B_{i'}^*$, denoted by $\hat{B}_{i'}^*$, can be written as 
$\hat{B}^*_{i'}=\epsilon\Gamma_{i',2^N-1}>0$ if $P_s>0$ (or $\hat{B}^*_{i'}=\epsilon\Lambda_{i',2^N-1}>0$ if $P_s=0$). Moreover, since $P_{i^{\dagger},2^N-1}^{*(1)}-(\epsilon\Gamma_{i',2^N-1}/\Gamma_{i^{\dagger},2^N-1})>0$ (or $P_{i^{\dagger},2^N-1}^{*(2)}-(\epsilon\Lambda_{i',2^N-1}/\Lambda_{i^{\dagger},2^N-1})>0$), the adjusted ${B}_{i^\dagger}^*$, i.e., $\hat{B}_{i^\dagger}^*$, satisfies $\hat{B}_{i^\dagger}^*>0$.

By following the same procedure for all $i'\in\mathcal{I}$ such that $A^*_{i'}>0$ and $B^*_{i'}=0$, we obtain a new solution based on the globally optimal solution, denoted by $\hat{\boldsymbol{\theta}}^{*}$, which gives rise to $\hat{A}^*_i>0$ and $\hat{B}^*_i>0$, $\forall i\in\mathcal{I}$. We know from (\ref{D1_function}) that this new solution $\hat{\boldsymbol{\theta}}^{*}$ leads to a finite objective value $D_1(\hat{\boldsymbol{\theta}}^{*})$, and hence $D_1(\hat{\boldsymbol{\theta}}^{*})<D_1({\boldsymbol{\theta}}^{*})$. This contradicts the optimality of the globally optimal solution $\boldsymbol{\theta}^*$ to (\ref{convex_KLD}). Therefore, the globally optimal solution to (\ref{convex_KLD}) cannot give rise to that $A^*_i>0$ and $B^*_i=0$ for any $i\in\mathcal{I}$.
\end{IEEEproof}
We can know from Claim \ref{claim_property} that for the globally optimal solution to (\ref{convex_KLD}), if $B^*_i=0$, then $A^*_i=0$, and if $A^*_i>0$, then $B^*_i>0$.
Next, we will solve (\ref{Appendix_q1*})--(\ref{Appendix_p2*}). First, we consider the optimization problem in (\ref{Appendix_q1*}). Since the problem in (\ref{convex_KLD}) is a convex optimization problem, the problem in (\ref{Appendix_q1*}) is also a convex optimization problem. Noting that the inequality constraints in (\ref{convex_KLD_ineq_q1}) of the problem in (\ref{Appendix_q1*}) are affine, and the feasible set of (\ref{Appendix_q1*}) is not empty, the weaker Slater’s condition for (\ref{Appendix_q1*}) holds, and hence, the Karush-Kuhn-Tucker (KKT) conditions are the necessary and sufficient conditions for the optimality of the solution to the problem in (\ref{Appendix_q1*}) \cite{boyd2004convex}. Hence, we can pursue the optimal solution to the optimization problem in  (\ref{Appendix_q1*}) by solving the KKT conditions for the problem in (\ref{Appendix_q1*}), which can be written as $\forall m\in\mathcal{M}_{\Delta}$,
\begin{subequations}\label{KKT_equations_q}
	\begin{align}
&\bigtriangledown_{{\textbf{Q}}_{m}^{(1)}}F_1({\textbf{Q}}_{m}^{(1)},\boldsymbol{\mu}_{1,m},\boldsymbol{\mu}_{2,m},\mu_{3,m})=0, \label{KKT_q_derivative_=0}\\
&\mu_{1,m}^{(i)}f_{{Q}_m^{(1)}}^{(i)}=0,\mu_{2,m}^{(i)}g_{{Q}_m^{(1)}}^{(i)}=0,\forall i\in\mathcal{I}, \label{KKT_muq_12_=0}\\   
&f_{{Q}_m^{(1)}}^{(i)},g_{{Q}_m^{(1)}}^{(i)}\le0,\forall i\in\mathcal{I},\label{KKT_fgq_1_<0}\\
&h_{{\textbf{Q}}_m^{(1)}}=0,\label{KKT_qh_1_=0}\\  
&\mu_{1,m}^{(i)},\mu_{2,m}^{(i)}\ge0,\forall i\in\mathcal{I},\label{KKT_muq_12_>0}
\end{align}
\end{subequations}
where $f_{ {Q}_m^{(1)}}^{(i)}\triangleq-  {Q}^{(1)}_{i,m}$, $g_{ {Q}_m^{(1)}}^{(i)}\triangleq{Q}^{(1)}_{i,m}-1$, $h_{{\textbf{Q}}_m^{(1)}}\triangleq\sum_{i\in\mathcal{I}}\Delta_{i,m}{Q}^{(1)}_{i,m}-P_s(1-\alpha)^{N-n_m}\alpha^{n_m}$, $\boldsymbol{\mu}_{1,m}$ is a vector stacking $\{{\mu}^{(i)}_{1,m}\}_{i\in\mathcal{I}}$, $\boldsymbol{\mu}_{2,m}$ is a vector stacking $\{{\mu}^{(i)}_{2,m}\}_{i\in\mathcal{I}}$, and
\begin{align}\label{equ_KKT_q_Lagrange}
\notag F_1({\textbf{Q}}_{m}^{(1)},\boldsymbol{\mu}_{1,m},\boldsymbol{\mu}_{2,m},\mu_{3,m})&\triangleq D^*_1\left({\textbf{Q}}_{m}^{(1)}\right)+\sum_{i\in\mathcal{I}}\mu_{1,m}^{(i)}f_{ {Q}_m^{(1)}}^{(i)}\\
&\quad+\sum_{i\in\mathcal{I}}\mu_{2,m}^{(i)} g_{ {Q}_m^{(1)}}^{(i)}+\mu_{3,m} h_{ \textbf{Q}_m^{(1)}}.\end{align}
Since for any $m\in\mathcal{M}_{\Delta}$ and any $i\in\mathcal{I}$, $f_{{Q}_m^{(1)}}^{(i)}=- {Q}_{i,m}^{(1)}\le0$, $g_{{Q}_m^{(1)}}^{(i)}={Q}_{i,m}^{(1)}-1\le0$, and $f_{{Q}_m^{(1)}}^{(i)}+g_{{Q}_m^{(1)}}^{(i)}=-1$, we know from (\ref{KKT_muq_12_=0}) that,
\begin{equation}\label{mu_condition2}
\mu_{1,m}^{(i)}\mu_{2,m}^{(i)}=0 ,\forall i\in\mathcal{I},\forall m\in\mathcal{M}_{\Delta}.
\end{equation}  
Let $\mu_{1,m}^{*(i)}$, $\mu_{2,m}^{*(i)}$, and $\mu_{3,m}^{*}$ be the optimal Lagrange multipliers.
For any $m\in\mathcal{M}_{\Delta}$ and any $ i\in\mathcal{I}$, we know from (\ref{KKT_muq_12_=0}), (\ref{KKT_fgq_1_<0}), (\ref{KKT_muq_12_>0}), and (\ref{mu_condition2}) that the optimal solution ${Q}_{i,m}^{*(1)}$ to (\ref{KKT_equations_q}) satisifies
\begin{numcases}{}
{Q}_{i,m}^{*(1)}=0,\text{ if }\mu_{1,m}^{*(i)}>0\text{ and }\mu_{2,m}^{*(i)}=0,\label{mu_condition2_2}\\
{Q}_{i,m}^{*(1)}=1,\text{ if }\mu_{1,m}^{*(i)}=0\text{ and }\mu_{2,m}^{*(i)}>0,\label{mu_condition2_3}\\
{Q}_{i,m}^{*(1)}\in[0,1],\text{ if }\mu_{1,m}^{*(i)}=\mu_{2,m}^{*(i)}=0.\label{mu_condition2_1}
\end{numcases}

For any $i\in\{i\in\mathcal{I}|B_i^*=0\}$, we know from Claim \ref{claim_property} that if $B_i^*=0$, then $A_i^*=0$. Moreover, we know from the definition of $A^*_i$ in Theorem \ref{Theorem_solution} that  $A_i^*=0$ implies that ${Q}_{i,m}^{*(1)}=0$ for all $m\in\mathcal{M}_{\Delta}$.

From (\ref{D1_function}) and (\ref{KKT_q_derivative_=0}), by taking the partial derivative of (\ref{equ_KKT_q_Lagrange}) with respect to ${Q}_{i,m}^{(1)}$ for any $m\in\mathcal{M}_{\Delta}$ and any $i\in\{i\in\mathcal{I}|B_i^*>0\}$ and setting it to zero, we can obtain that $\forall m\in\mathcal{M}_{\Delta}$ and $\forall i\in\{i\in\mathcal{I}|B_i^*>0\}$, the optimal solution ${{Q}}_{i,m}^{*(1)}$ satisfies
\begin{align}\label{derivative_q_ori} \notag
&\ln\frac{ \!\Theta_i\!\!+\!\!\!\sum\limits_{m'\in\mathcal{M}_{\Delta}\backslash\{m\}}\!\!\Delta_{i,m'}{Q}^{*(1)}_{i,m'}\!+\!\Delta_{i,m}{Q}^{*(1)}_{i,m}\!\!+\!\!\!\!\sum\limits_{m'\in\mathcal{M}_{\Phi}}\!\!\Phi_{i,m'}{Q}^{*(2)}_{i,m'}\!}{\Psi_i+\sum\limits_{m'\in\mathcal{M}_{\Gamma}}\Gamma_{i,m'}{P}^{*(1)}_{i,m'}+\sum\limits_{m'\in\mathcal{M}_{\Lambda}}\Lambda_{i,m'}{P}^{*(2)}_{i,m'}}\\
&\times \Delta_{i,m}+\Delta_{i,m}-\mu_{1,m}^{*(i)}+\mu_{2,m}^{*(i)}+\mu_{3,m}^{*}\Delta_{i,m}=0,\end{align}
which is equivalent to
\begin{align}\label{derivative_q} \notag
&\ln\frac{ \!\Theta_i\!\!+\!\!\!\sum\limits_{m'\in\mathcal{M}_{\Delta}\backslash\{m\}}\!\!\Delta_{i,m'}{Q}^{*(1)}_{i,m'}\!+\!\Delta_{i,m}{Q}^{*(1)}_{i,m}\!\!+\!\!\!\!\sum\limits_{m'\in\mathcal{M}_{\Phi}}\!\!\Phi_{i,m'}{Q}^{*(2)}_{i,m'}\!}{\Psi_i+\sum\limits_{m'\in\mathcal{M}_{\Gamma}}\Gamma_{i,m'}{P}^{*(1)}_{i,m'}+\sum\limits_{m'\in\mathcal{M}_{\Lambda}}\Lambda_{i,m'}{P}^{*(2)}_{i,m'}}\\
&=\frac{-\Delta_{i,m}+\mu_{1,m}^{*(i)}-\mu_{2,m}^{*(i)}-\mu_{3,m}^{*}\Delta_{i,m}}{\Delta_{i,m}}.\end{align}

Next, we will develop ${Q}_{i,m}^{*(1)}$ for any $m\in\mathcal{M}_{\Delta}$, any $i\in\{i\in\mathcal{I}|B_i^*>0\}$, and different value of $\mu_{3,m}^{*}$ by using (\ref{derivative_q}).
We first define 
\begin{align}\label{equ_gamma_q}	
\zeta_{\textbf{q},m}^{*(1)}\triangleq \text{exp}\left\{-\mu_{3,m}^{*}-1\right\},
\end{align}
which only depends on $\mu_{3,m}^{*}$. 

\begin{claim}\label{claim_q_<}For any $m\in\mathcal{M}_{\Delta}$ and any $ i\in\{i\in\mathcal{I}|B_i^*>0\}$, if $\zeta_{\textbf{q},m}^{*(1)}B^*_i\le \Theta_i+\sum_{m'\in\mathcal{M}_{\Delta}\backslash\{m\}}\Delta_{i,m'}{Q}^{*(1)}_{i,m'}+\sum_{m'\in\mathcal{M}_{\Phi}}\Phi_{i,m'}{Q}^{*(2)}_{i,m'}$, then in order to make (\ref{derivative_q}) hold, ${Q}_{i,m}^{*(1)}=0$.
\end{claim}
\begin{IEEEproof}[Proof of Claim \ref{claim_q_<}]
For any $m\in\mathcal{M}_{\Delta}$ and any $ i\in\{i\in\mathcal{I}|B_i^*>0\}$, if $\zeta_{\textbf{q},m}^{*(1)}B^*_i\le \Theta_i+\sum_{m'\in\mathcal{M}_{\Delta}\backslash\{m\}}\Delta_{i,m'}{Q}^{*(1)}_{i,m'}+\sum_{m'\in\mathcal{M}_{\Phi}}\Phi_{i,m'}{Q}^{*(2)}_{i,m'}$, then by using the definition of $B^*_i$ in Theorem \ref{Theorem_solution}, we can obtain
\begin{align}\label{equ_q_case_1_1}
\frac{ \Theta_i\!\!+\!\!\!\!\sum\limits_{m'\in\mathcal{M}_{\Delta}\backslash\{m\}}\!\!\Delta_{i,m'}{Q}^{*(1)}_{i,m'}\!\!+\!\!\sum\limits_{m'\in\mathcal{M}_{\Phi}}\!\!\Phi_{i,m'}{Q}^{*(2)}_{i,m'}}{\Psi_i+\sum\limits_{m'\in\mathcal{M}_{\Gamma}}\Gamma_{i,m'}{P}^{*(1)}_{i,m'}+\sum\limits_{m'\in\mathcal{M}_{\Lambda}}\Lambda_{i,m'}{P}^{*(2)}_{i,m'}}\!\ge\!\zeta_{\textbf{q},m}^{*(1)}.
\end{align}
Noting that $\Delta_{i,m}> 0$ and ${Q}_{i,m}^{*(1)}\ge 0$, we can obtain from (\ref{equ_q_case_1_1}) that 
\begin{align}\label{equ_q_case_1_2}\notag
&\frac{\! \Theta_i\!+\!\!\sum\limits_{m'\in\mathcal{M}_{\Delta}\backslash\{m\}}\!\!\Delta_{i,m'}{Q}^{*(1)}_{i,m'}+\Delta_{i,m}{Q}^{*(1)}_{i,m}\!+\!\!\sum\limits_{m'\in\mathcal{M}_{\Phi}}\!\!\Phi_{i,m'}{Q}^{*(2)}_{i,m'}}{\Psi_i+\sum\limits_{m'\in\mathcal{M}_{\Gamma}}\Gamma_{i,m'}{P}^{*(1)}_{i,m'}+\sum\limits_{m'\in\mathcal{M}_{\Lambda}}\Lambda_{i,m'}{P}^{*(2)}_{i,m'}}\\ \notag
&\ge\frac{ \Theta_i+\sum\limits_{m'\in\mathcal{M}_{\Delta}\backslash\{m\}}\Delta_{i,m'}{Q}^{*(1)}_{i,m'}+\sum\limits_{m'\in\mathcal{M}_{\Phi}}\Phi_{i,m'}{Q}^{*(2)}_{i,m'}}{\Psi_i+\sum\limits_{m'\in\mathcal{M}_{\Gamma}}\Gamma_{i,m'}{P}^{*(1)}_{i,m'}+\sum\limits_{m'\in\mathcal{M}_{\Lambda}}\Lambda_{i,m'}{P}^{*(2)}_{i,m'}}\\
&\ge\zeta_{\textbf{q},m}^{*(1)}.
\end{align}
Comparing (\ref{equ_q_case_1_2}) with (\ref{derivative_q}), we can obtain that in order to make (\ref{derivative_q}) hold, 
\begin{align}\label{equ_q_case_1_4}
\frac{ \mu_{1,m}^{*(i)}-\mu_{2,m}^{*(i)}}{\Delta_{i,m}}\ge 0,
\end{align}
which yields $\mu_{1,m}^{*(i)}\ge0$ and $\mu_{2,m}^{*(i)}=0$ due to the fact that there are only three combinations of the signs of $\mu_{1,m}^{*(i)}$ and $\mu_{2,m}^{*(i)}$ which are illustrated in (\ref{mu_condition2_2}), (\ref{mu_condition2_3}), and (\ref{mu_condition2_1}).
If $\mu_{1,m}^{*(i)}>0$ and $\mu_{2,m}^{*(i)}=0$, then from (\ref{mu_condition2_2}), we can obtain ${Q}_{i,m}^{*(1)}=0$. If $\mu_{1,m}^{*(i)}=0$ and $\mu_{2,m}^{*(i)}=0$, then (\ref{derivative_q}) is equivalent to
\begin{align}\label{equ_q_case_1_5}\notag
&\frac{ \!\Theta_i\!+\!\sum\limits_{m'\in\mathcal{M}_{\Delta}\backslash\{m\}}\!\!\Delta_{i,m'}{Q}^{*(1)}_{i,m'}\!+\!\Delta_{i,m}{Q}^{*(1)}_{i,m}\!+\!\sum\limits_{m'\in\mathcal{M}_{\Phi}}\!\!\Phi_{i,m'}{Q}^{*(2)}_{i,m'}}{\Psi_i+\sum\limits_{m'\in\mathcal{M}_{\Gamma}}\Gamma_{i,m'}{P}^{*(1)}_{i,m'}+\sum\limits_{m'\in\mathcal{M}_{\Lambda}}\Lambda_{i,m'}{P}^{*(2)}_{i,m'}}\\
&=\zeta_{\textbf{q},m}^{*(1)}.\end{align}
From (\ref{equ_q_case_1_2}) and (\ref{equ_q_case_1_5}), we have
\begin{align}\label{equ_q_case_1_6}\notag
&\frac{ \!\Theta_i\!+\!\sum\limits_{m'\in\mathcal{M}_{\Delta}\backslash\{m\}}\!\!\Delta_{i,m'}{Q}^{*(1)}_{i,m'}\!+\!\Delta_{i,m}{Q}^{*(1)}_{i,m}\!+\!\sum\limits_{m'\in\mathcal{M}_{\Phi}}\!\!\Phi_{i,m'}{Q}^{*(2)}_{i,m'}}{\Psi_i+\sum\limits_{m'\in\mathcal{M}_{\Gamma}}\Gamma_{i,m'}{P}^{*(1)}_{i,m'}+\sum\limits_{m'\in\mathcal{M}_{\Lambda}}\Lambda_{i,m'}{P}^{*(2)}_{i,m'}}\\
&=\frac{ \Theta_i+\sum\limits_{m'\in\mathcal{M}_{\Delta}\backslash\{m\}}\Delta_{i,m'}{Q}^{*(1)}_{i,m'}+\sum\limits_{m'\in\mathcal{M}_{\Phi}}\Phi_{i,m'}{Q}^{*(2)}_{i,m'}}{\Psi_i+\sum\limits_{m'\in\mathcal{M}_{\Gamma}}\Gamma_{i,m'}{P}^{*(1)}_{i,m'}+\sum\limits_{m'\in\mathcal{M}_{\Lambda}}\Lambda_{i,m'}{P}^{*(2)}_{i,m'}},\end{align}
which implies that ${Q}_{i,m}^{*(1)}=0$. The proof of Claim \ref{claim_q_<} is completed.
\end{IEEEproof}

\begin{claim}\label{claim_q_>}For any $m\in\mathcal{M}_{\Delta}$ and any $ i\in\{i\in\mathcal{I}|B_i^*>0\}$, if $\zeta_{\textbf{q},m}^{*(1)}B^*_i\ge \Theta_i+\sum_{m'\in\mathcal{M}_{\Delta}\backslash\{m\}}\Delta_{i,m'}{Q}^{*(1)}_{i,m'}+\Delta_{i,m}+\sum_{m'\in\mathcal{M}_{\Phi}}\Phi_{i,m'}{Q}^{*(2)}_{i,m'}$, then in order to make (\ref{derivative_q}) hold, ${Q}_{i,m}^{*(1)}=1$.
\end{claim}
\begin{IEEEproof}[Proof of Claim \ref{claim_q_>}]
For any $m\in\mathcal{M}_{\Delta}$ and any $ i\in\{i\in\mathcal{I}|B_i^*>0\}$, if $\zeta_{\textbf{q},m}^{*(1)}B^*_i\ge \Theta_i+\sum_{m'\in\mathcal{M}_{\Delta}\backslash\{m\}}\Delta_{i,m'}{Q}^{*(1)}_{i,m'}+\Delta_{i,m}+\sum_{m'\in\mathcal{M}_{\Phi}}\Phi_{i,m'}{Q}^{*(2)}_{i,m'}$,  then by using the definition of $B^*_i$ in Theorem \ref{Theorem_solution}, we can obtain
\begin{align}\label{equ_q_case_2_1}\notag
&\frac{ \Theta_i+\sum\limits_{m'\in\mathcal{M}_{\Delta}\backslash\{m\}}\Delta_{i,m'}{Q}^{*(1)}_{i,m'}+\Delta_{i,m}+\sum\limits_{m'\in\mathcal{M}_{\Phi}}\Phi_{i,m'}{Q}^{*(2)}_{i,m'}}{\Psi_i+\sum\limits_{m'\in\mathcal{M}_{\Gamma}}\Gamma_{i,m'}{P}^{*(1)}_{i,m'}+\sum\limits_{m'\in\mathcal{M}_{\Lambda}}\Lambda_{i,m'}{P}^{*(2)}_{i,m'}}\\
&\le\zeta_{\textbf{q},m}^{*(1)}.
\end{align}
Noting that $\Delta_{i,m}> 0$ and ${Q}_{i,m}^{*(1)}\le 1$, we can obtain from (\ref{equ_q_case_2_1}) that 
\begin{align}\label{equ_q_case_2_2}\notag
&\frac{ \!\Theta_i\!+\!\sum\limits_{m'\in\mathcal{M}_{\Delta}\backslash\{m\}}\!\!\Delta_{i,m'}{Q}^{*(1)}_{i,m'}\!+\!\Delta_{i,m}{Q}^{*(1)}_{i,m}\!+\!\!\sum\limits_{m'\in\mathcal{M}_{\Phi}}\!\!\Phi_{i,m'}{Q}^{*(2)}_{i,m'}}{\Psi_i+\sum\limits_{m'\in\mathcal{M}_{\Gamma}}\Gamma_{i,m'}{P}^{*(1)}_{i,m'}+\sum\limits_{m'\in\mathcal{M}_{\Lambda}}\Lambda_{i,m'}{P}^{*(2)}_{i,m'}}\\ \notag
&\le\frac{\! \Theta_i\!+\!\!\sum\limits_{m'\in\mathcal{M}_{\Delta}\backslash\{m\}}\!\!\Delta_{i,m'}{Q}^{*(1)}_{i,m'}\!+\!\Delta_{i,m}\!+\!\!\sum\limits_{m'\in\mathcal{M}_{\Phi}}\!\!\Phi_{i,m'}{Q}^{*(2)}_{i,m'}}{\Psi_i+\sum\limits_{m'\in\mathcal{M}_{\Gamma}}\Gamma_{i,m'}{P}^{*(1)}_{i,m'}+\sum\limits_{m'\in\mathcal{M}_{\Lambda}}\Lambda_{i,m'}{P}^{*(2)}_{i,m'}}\\
&\le\zeta_{\textbf{q},m}^{*(1)}.
\end{align}
Comparing (\ref{equ_q_case_2_2}) with (\ref{derivative_q}), we can obtain that in order to make (\ref{derivative_q}) hold, 
\begin{align}\label{equ_q_case_2_4}
\frac{ \mu_{2,m}^{*(i)}-\mu_{1,m}^{*(i)}}{\Delta_{i,m}}\ge 0,
\end{align}
which yields $\mu_{1,m}^{*(i)}=0$ and $\mu_{2,m}^{*(i)}\ge0$ due to the fact that there are only three combinations of the signs of $\mu_{1,m}^{*(i)}$ and $\mu_{2,m}^{*(i)}$ which are illustrated in (\ref{mu_condition2_2}), (\ref{mu_condition2_3}), and (\ref{mu_condition2_1}).
If $\mu_{1,m}^{*(i)}=0$ and $\mu_{2,m}^{*(i)}>0$, then from (\ref{mu_condition2_3}), we can obtain ${Q}_{i,m}^{*(1)}=1$. If $\mu_{1,m}^{*(i)}=0$ and $\mu_{2,m}^{*(i)}=0$, then (\ref{derivative_q}) is equivalent to
\begin{align}\label{equ_q_case_2_5}\notag
&\frac{\!\Theta_i\!+\!\sum\limits_{m'\in\mathcal{M}_{\Delta}\backslash\{m\}}\!\Delta_{i,m'}{Q}^{*(1)}_{i,m'}\!+\!\!\Delta_{i,m}{Q}^{*(1)}_{i,m}\!+\!\!\sum\limits_{m'\in\mathcal{M}_{\Phi}}\!\!\Phi_{i,m'}{Q}^{*(2)}_{i,m'}}{\Psi_i+\sum\limits_{m'\in\mathcal{M}_{\Gamma}}\Gamma_{i,m'}{P}^{*(1)}_{i,m'}+\sum\limits_{m'\in\mathcal{M}_{\Lambda}}\Lambda_{i,m'}{P}^{*(2)}_{i,m'}}\\
&=\zeta_{\textbf{q},m}^{*(1)}.\end{align}
From (\ref{equ_q_case_2_2}) and (\ref{equ_q_case_2_5}), we have
\begin{align}\label{equ_q_case_2_6}\notag
&\frac{ \!\Theta_i\!+\!\!\!\sum\limits_{m'\in\mathcal{M}_{\Delta}\backslash\{m\}}\!\!\Delta_{i,m'}{Q}^{*(1)}_{i,m'}\!+\!\!\Delta_{i,m}{Q}^{*(1)}_{i,m}\!+\!\!\sum\limits_{m'\in\mathcal{M}_{\Phi}}\!\!\Phi_{i,m'}{Q}^{*(2)}_{i,m'}}{\Psi_i+\sum\limits_{m'\in\mathcal{M}_{\Gamma}}\Gamma_{i,m'}{P}^{*(1)}_{i,m'}+\sum\limits_{m'\in\mathcal{M}_{\Lambda}}\Lambda_{i,m'}{P}^{*(2)}_{i,m'}}\\ 
&=\!\!\frac{ \!\Theta_i\!\!+\!\!\!\sum\limits_{m'\in\mathcal{M}_{\Delta}\backslash\{m\}}\!\!\!\Delta_{i,m'}{Q}^{*(1)}_{i,m'}\!+\!\!\Delta_{i,m}\!\!+\!\!\!\!\sum\limits_{m'\in\mathcal{M}_{\Phi}}\!\!\!\!\Phi_{i,m'}{Q}^{*(2)}_{i,m'}}{\Psi_i+\sum\limits_{m'\in\mathcal{M}_{\Gamma}}\Gamma_{i,m'}{P}^{*(1)}_{i,m'}+\sum\limits_{m'\in\mathcal{M}_{\Lambda}}\Lambda_{i,m'}{P}^{*(2)}_{i,m'}},\end{align}
which implies that ${Q}_{i,m}^{*(1)}=1$. The proof of Claim \ref{claim_q_>} is completed.
\end{IEEEproof}

\begin{claim}\label{claim_q_=}For any $m\in\mathcal{M}_{\Delta}$ and any $ i\in\{i\in\mathcal{I}|B_i^*>0\}$, if $\zeta_{\textbf{q},m}^{*(1)}B^*_i\in (\Theta_i+\sum_{m'\in\mathcal{M}_{\Delta}\backslash\{m\}}\Delta_{i,m'}{Q}^{*(1)}_{i,m'}+\sum_{m'\in\mathcal{M}_{\Phi}}\Phi_{i,m'}{Q}^{*(2)}_{i,m'},\Theta_i+\sum_{m'\in\mathcal{M}_{\Delta}\backslash\{m\}}\Delta_{i,m'}{Q}^{*(1)}_{i,m'}+\Delta_{i,m}+\sum_{m'\in\mathcal{M}_{\Phi}}\Phi_{i,m'}{Q}^{*(2)}_{i,m'})$, then in order to make (\ref{derivative_q}) hold, ${Q}_{i,m}^{*(1)}=(\zeta_{\textbf{q},m}^{*(1)}B
^*_{i}-\Theta_i-\sum_{m'\in\mathcal{M}_{\Delta}\backslash\{m\}}\Delta_{i,m'}{Q}^{*(1)}_{i,m'}-\sum_{m'\in\mathcal{M}_{\Phi}}\Phi_{i,m'}{Q}^{*(2)}_{i,m'})/\Delta_{i,m}$.
\end{claim}
\begin{IEEEproof}[Proof of Claim \ref{claim_q_=}]
First, we show that for any $m\in\mathcal{M}_{\Delta}$ and any $ i\in\{i\in\mathcal{I}|B_i^*>0\}$, in order to make (\ref{derivative_q}) hold,  $\mu_{1,m}^{*(i)}=\mu_{2,m}^{*(i)}=0$ for the case that $\zeta_{\textbf{q},m}^{*(1)}B^*_i\in (\Theta_i+\sum_{m'\in\mathcal{M}_{\Delta}\backslash\{m\}}\Delta_{i,m'}{Q}^{*(1)}_{i,m'}+\sum_{m'\in\mathcal{M}_{\Phi}}\Phi_{i,m'}{Q}^{*(2)}_{i,m'},\Theta_i+\sum_{m'\in\mathcal{M}_{\Delta}\backslash\{m\}}\Delta_{i,m'}{Q}^{*(1)}_{i,m'}+\Delta_{i,m}+\sum_{m'\in\mathcal{M}_{\Phi}}\Phi_{i,m'}{Q}^{*(2)}_{i,m'})$.

Note that there are only three combinations of the signs of $\mu_{1,m}^{*(i)}$ and $\mu_{2,m}^{*(i)}$ which are illustrated in (\ref{mu_condition2_2}), (\ref{mu_condition2_3}), and (\ref{mu_condition2_1}). If  $\mu_{1,m}^{*(i)}>0$ and  $\mu_{2,m}^{*(i)}=0$, then from (\ref{mu_condition2_2}) we know that ${Q}_{i,m}^{*(1)}=0$.  Hence, (\ref{derivative_q}) is equivalent to
\begin{align}\label{equ_q_case_3_1}\notag
&\ln\frac{ \Theta_i+\sum\limits_{m'\in\mathcal{M}_{\Delta}\backslash\{m\}}\Delta_{i,m'}{Q}^{*(1)}_{i,m'}+\sum\limits_{m'\in\mathcal{M}_{\Phi}}\Phi_{i,m'}{Q}^{*(2)}_{i,m'}}{\Psi_i+\sum\limits_{m'\in\mathcal{M}_{\Gamma}}\Gamma_{i,m'}{P}^{*(1)}_{i,m'}+\sum\limits_{m'\in\mathcal{M}_{\Lambda}}\Lambda_{i,m'}{P}^{*(2)}_{i,m'}}\\
&=\frac{-\Delta_{i,m}+\mu_{1,m}^{*(i)}-\mu_{3,m}^{*}\Delta_{i,m}}{\Delta_{i,m}},\end{align}
which yields 
\begin{align}\label{equ_q_case_3_2}\notag
&\frac{ \Theta_i+\sum\limits_{m'\in\mathcal{M}_{\Delta}\backslash\{m\}}\Delta_{i,m'}{Q}^{*(1)}_{i,m'}+\sum\limits_{m'\in\mathcal{M}_{\Phi}}\Phi_{i,m'}{Q}^{*(2)}_{i,m'}}{\Psi_i+\sum\limits_{m'\in\mathcal{M}_{\Gamma}}\Gamma_{i,m'}{P}^{*(1)}_{i,m'}+\sum\limits_{m'\in\mathcal{M}_{\Lambda}}\Lambda_{i,m'}{P}^{*(2)}_{i,m'}}\\
&>\zeta_{\textbf{q},m}^{*(1)}.\end{align}
However, for the case that $\zeta_{\textbf{q},m}^{*(1)}B^*_i\in (\Theta_i+\sum_{m'\in\mathcal{M}_{\Delta}\backslash\{m\}}\Delta_{i,m'}{Q}^{*(1)}_{i,m'}+\sum_{m'\in\mathcal{M}_{\Phi}}\Phi_{i,m'}{Q}^{*(2)}_{i,m'},\Theta_i+\sum_{m'\in\mathcal{M}_{\Delta}\backslash\{m\}}\Delta_{i,m'}{Q}^{*(1)}_{i,m'}+\Delta_{i,m}+\sum_{m'\in\mathcal{M}_{\Phi}}\Phi_{i,m'}{Q}^{*(2)}_{i,m'})$, we have $\zeta_{\textbf{q},m}^{*(1)}B^*_i>\Theta_i+\sum_{m'\in\mathcal{M}_{\Delta}\backslash\{m\}}\Delta_{i,m'}{Q}^{*(1)}_{i,m'}+\sum_{m'\in\mathcal{M}_{\Phi}}\Phi_{i,m'}{Q}^{*(2)}_{i,m'}$ which implies that 
\begin{align}\label{equ_q_case_3_2_1} \notag 
&\frac{ \Theta_i+\sum\limits_{m'\in\mathcal{M}_{\Delta}\backslash\{m\}}\Delta_{i,m'}{Q}^{*(1)}_{i,m'}\!+\!\sum\limits_{m'\in\mathcal{M}_{\Phi}}\!\Phi_{i,m'}{Q}^{*(2)}_{i,m'}}{\Psi_i+\sum\limits_{m'\in\mathcal{M}_{\Gamma}}\Gamma_{i,m'}{P}^{*(1)}_{i,m'}+\sum\limits_{m'\in\mathcal{M}_{\Lambda}}\Lambda_{i,m'}{P}^{*(2)}_{i,m'}}\\
&<\zeta_{\textbf{q},m}^{*(1)},\end{align}
which contradicts (\ref{equ_q_case_3_2}). Hence, $\mu_{1,m}^{*(i)}>0$ and  $\mu_{2,m}^{*(i)}=0$ cannot make (\ref{derivative_q}) hold for this case. If  $\mu_{1,m}^{*(i)}=0$ and  $\mu_{2,m}^{*(i)}>0$, then from (\ref{mu_condition2_3}) we know that ${Q}_{i,m}^{*(1)}=1$.  Hence, (\ref{derivative_q}) is equivalent to
\begin{align}\label{equ_q_case_3_3}\notag
&\ln\frac{ \!\Theta_i\!+\!\sum\limits_{m'\in\mathcal{M}_{\Delta}\backslash\{m\}}\!\!\Delta_{i,m'}{Q}^{*(1)}_{i,m'}\!+\!\Delta_{i,m}\!+\!\sum\limits_{m'\in\mathcal{M}_{\Phi}}\!\!\Phi_{i,m'}{Q}^{*(2)}_{i,m'}}{\Psi_i+\sum\limits_{m'\in\mathcal{M}_{\Gamma}}\Gamma_{i,m'}{P}^{*(1)}_{i,m'}+\sum\limits_{m'\in\mathcal{M}_{\Lambda}}\Lambda_{i,m'}{P}^{*(2)}_{i,m'}}\\
&=\frac{-\Delta_{i,m}-\mu_{2,m}^{*(i)}-\mu_{3,m}^{*}\Delta_{i,m}}{\Delta_{i,m}},\end{align}
which yields 
\begin{align}\label{equ_q_case_3_4}\notag
&\frac{ \Theta_i+\sum\limits_{m'\in\mathcal{M}_{\Delta}\backslash\{m\}}\Delta_{i,m'}{Q}^{*(1)}_{i,m'}+\Delta_{i,m}+\sum\limits_{m'\in\mathcal{M}_{\Phi}}\Phi_{i,m'}{Q}^{*(2)}_{i,m'}}{\Psi_i+\sum\limits_{m'\in\mathcal{M}_{\Gamma}}\Gamma_{i,m'}{P}^{*(1)}_{i,m'}+\sum\limits_{m'\in\mathcal{M}_{\Lambda}}\Lambda_{i,m'}{P}^{*(2)}_{i,m'}}\\
&<\zeta_{\textbf{q},m}^{*(1)}.\end{align}
However, the condition $\zeta_{\textbf{q},m}^{*(1)}B^*_i<\Theta_i+\sum_{m'\in\mathcal{M}_{\Delta}\backslash\{m\}}\Delta_{i,m'}{Q}^{*(1)}_{i,m'}+\Delta_{i,m}+\sum_{m'\in\mathcal{M}_{\Phi}}\Phi_{i,m'}{Q}^{*(2)}_{i,m'}$ implies that 
\begin{align}\label{equ_q_case_3_4_1}\notag
&\frac{ \Theta_i+\sum\limits_{m'\in\mathcal{M}_{\Delta}\backslash\{m\}}\Delta_{i,m'}{Q}^{*(1)}_{i,m'}+\Delta_{i,m}+\sum\limits_{m'\in\mathcal{M}_{\Phi}}\Phi_{i,m'}{Q}^{*(2)}_{i,m'}}{\Psi_i+\sum\limits_{m'\in\mathcal{M}_{\Gamma}}\Gamma_{i,m'}{P}^{*(1)}_{i,m'}+\sum\limits_{m'\in\mathcal{M}_{\Lambda}}\Lambda_{i,m'}{P}^{*(2)}_{i,m'}}\\
&>\zeta_{\textbf{q},m}^{*(1)},\end{align}
which contradicts (\ref{equ_q_case_3_4}). Hence, $\mu_{1,m}^{*(i)}=0$ and  $\mu_{2,m}^{*(i)}>0$ cannot make (\ref{derivative_q}) hold for this case.  For the last case that $\mu_{1,m}^{*(i)}=0$ and  $\mu_{2,m}^{*(i)}=0$, we know from (\ref{mu_condition2_1}) that ${Q}_{i,m}^{*(1)}\in[0,1]$, and (\ref{derivative_q}) is equivalent to
\begin{align}\label{equ_q_case_3_5}\notag
&\ln\frac{ \!\Theta_i\!+\!\!\!\sum\limits_{m'\in\mathcal{M}_{\Delta}\backslash\{m\}}\!\!\Delta_{i,m'}{Q}^{*(1)}_{i,m'}\!\!+\!\!\Delta_{i,m}{Q}_{i,m}^{*(1)}\!\!+\!\!\!\sum\limits_{m'\in\mathcal{M}_{\Phi}}\!\!\Phi_{i,m'}{Q}^{*(2)}_{i,m'}\!}{\Psi_i+\sum\limits_{m'\in\mathcal{M}_{\Gamma}}\Gamma_{i,m'}{P}^{*(1)}_{i,m'}+\sum\limits_{m'\in\mathcal{M}_{\Lambda}}\Lambda_{i,m'}{P}^{*(2)}_{i,m'}}\\
&=\frac{-\Delta_{i,m}-\mu_{3,m}^{*}\Delta_{i,m}}{\Delta_{i,m}},\end{align}
which yields 
\begin{align}\label{equ_q_case_3_6}
{Q}_{i,m}^{*(1)}=\frac{\left({\zeta_{\textbf{q},m}^{*(1)}B^*_{i}-\Theta_i-\sum_{m'\in\mathcal{M}_{\Phi}}\Phi_{i,m'}{Q}^{*(2)}_{i,m'}}\atop-\sum_{m'\in\mathcal{M}_{\Delta}\backslash\{m\}}\Delta_{i,m'}{Q}^{*(1)}_{i,m'}\right)}{\Delta_{i,m}}.\end{align}
Since $\zeta_{\textbf{q},m}^{*(1)}B^*_i\in (\Theta_i+\sum_{m'\in\mathcal{M}_{\Delta}\backslash\{m\}}\Delta_{i,m'}{Q}^{*(1)}_{i,m'}+\sum_{m'\in\mathcal{M}_{\Phi}}\Phi_{i,m'}{Q}^{*(2)}_{i,m'},\Theta_i+\sum_{m'\in\mathcal{M}_{\Delta}\backslash\{m\}}\Delta_{i,m'}{Q}^{*(1)}_{i,m'}+\Delta_{i,m}+\sum_{m'\in\mathcal{M}_{\Phi}}\Phi_{i,m'}{Q}^{*(2)}_{i,m'})$, we can know from (\ref{equ_q_case_3_6}) that ${Q}_{i,m}^{*(1)}\in(0,1)$ which  satisfies the constraint ${Q}_{i,m}^{*(1)}\in[0,1]$. Therefore,  for any $m\in\mathcal{M}_{\Delta}$ and any $ i\in\{i\in\mathcal{I}|B_i^*>0\}$, in order to make (\ref{derivative_q}) hold,  $\mu_{1,m}^{*(i)}=\mu_{2,m}^{*(i)}=0$ if $\zeta_{\textbf{q},m}^{*(1)}B^*_i\in (\Theta_i+\sum_{m'\in\mathcal{M}_{\Delta}\backslash\{m\}}\Delta_{i,m'}{Q}^{*(1)}_{i,m'}+\sum_{m'\in\mathcal{M}_{\Phi}}\Phi_{i,m'}{Q}^{*(2)}_{i,m'},\Theta_i+\sum_{m'\in\mathcal{M}_{\Delta}\backslash\{m\}}\Delta_{i,m'}{Q}^{*(1)}_{i,m'}+\Delta_{i,m}+\sum_{m'\in\mathcal{M}_{\Phi}}\Phi_{i,m'}{Q}^{*(2)}_{i,m'})$. From (\ref{derivative_q}) and $\mu_{1,m}^{*(i)}=\mu_{2,m}^{*(i)}=0$, we can get (\ref{equ_q_case_3_6}) which completes the proof of Claim \ref{claim_q_=}.
\end{IEEEproof}
Note that (\ref{Appendix_q1*}) holds for any $m\in\mathcal{M}_{\Delta}$. Hence from Claim \ref{claim_property}, \ref{claim_q_<},  \ref{claim_q_>}, and  \ref{claim_q_=}, the expression for ${Q}_{i,m}^{*(1)}$ can be summarized as that $\forall i\in\mathcal{I}$ and $\forall m\in\mathcal{M}_{\Delta}$,
\begin{align}\label{equ_Q_1} 
{Q}_{i,m}^{*(1)}\!\!=\!\!\min\!\left\{\!\!\left[\!\frac{\left({\zeta_{\textbf{q},m}^{*(1)}B^*_{i}\!-\!\Theta_i\!-\!\sum_{m'\in\mathcal{M}_{\Phi}}\!\!\Phi_{i,m'}{Q}^{*(2)}_{i,m'}}\atop-\sum_{m'\in\mathcal{M}_{\Delta}\backslash\{m\}}\Delta_{i,m'}{Q}^{*(1)}_{i,m'}\right)}{\Delta_{i,m}}\!\right]^+\!\!,\!1\!\right\}\!.
\end{align}
Note that (\ref{equ_Q_1}) only satisfies the constraint ${Q}_{i,m}^{*(1)}\in[0,1]$. In order to satisfy the constraint in (\ref{KKT_qh_1_=0}), $\zeta_{\textbf{q},m}^{*(1)}$ should ensure $\sum_{i\in\mathcal{I}}\Delta_{i,m}{Q}_{i,m}^{*(1)}=P_s(1-\alpha)^{N-n_m}\alpha^{n_m}$. It's worth mentioning that we know from (\ref{equ_Q_1}) that $\zeta_{\textbf{q},m}^{*(1)}$ must be positive so that the constraint in (\ref{KKT_qh_1_=0}) can be satisfied. 

The processes of solving (\ref{Appendix_q2*}), (\ref{Appendix_p1*}), and (\ref{Appendix_p2*}) are similar to the process of solving (\ref{Appendix_q1*}). Similar to (\ref{equ_Q_1}), the solutions ${Q}_{i,m,}^{*(2)}$, ${P}_{i,m,}^{*(1)}$, and ${P}_{i,m,}^{*(2)}$ to the optimization problems in (\ref{Appendix_q2*}), (\ref{Appendix_p1*}), and (\ref{Appendix_p2*}) can be expressed as (\ref{Thm_qb2}), (\ref{Thm_pb1}), and (\ref{Thm_pb2}), respectively, which completes the proof.
\ifCLASSOPTIONcaptionsoff
  \newpage
\fi

\bibliographystyle{IEEEtran}
\bibliography{IEEEabrv,Blockchain}

\begin{thebibliography}{10}
\providecommand{\url}[1]{#1}
\csname url@samestyle\endcsname
\providecommand{\newblock}{\relax}
\providecommand{\bibinfo}[2]{#2}
\providecommand{\BIBentrySTDinterwordspacing}{\spaceskip=0pt\relax}
\providecommand{\BIBentryALTinterwordstretchfactor}{4}
\providecommand{\BIBentryALTinterwordspacing}{\spaceskip=\fontdimen2\font plus
\BIBentryALTinterwordstretchfactor\fontdimen3\font minus
  \fontdimen4\font\relax}
\providecommand{\BIBforeignlanguage}[2]{{%
\expandafter\ifx\csname l@#1\endcsname\relax
\typeout{** WARNING: IEEEtran.bst: No hyphenation pattern has been}%
\typeout{** loaded for the language `#1'. Using the pattern for}%
\typeout{** the default language instead.}%
\else
\language=\csname l@#1\endcsname
\fi
#2}}
\providecommand{\BIBdecl}{\relax}
\BIBdecl

\bibitem{al2015internet}
A.~Al-Fuqaha, M.~Guizani, M.~Mohammadi, M.~Aledhari, and M.~Ayyash, ``{Internet
  of Things}: A survey on enabling technologies, protocols, and applications,''
  \emph{IEEE Commun. Surveys Tuts.}, vol.~17, no.~4, pp. 2347--2376, 2015.

\bibitem{lin2017survey}
J.~Lin, W.~Yu, N.~Zhang, X.~Yang, H.~Zhang, and W.~Zhao, ``A survey on
  {Internet of Things}: Architecture, enabling technologies, security and
  privacy, and applications,'' \emph{IEEE Internet Things J.}, vol.~4, no.~5,
  pp. 1125--1142, 2017.

\bibitem{mosenia2016comprehensive}
A.~Mosenia and N.~K. Jha, ``A comprehensive study of security of
  {Internet-of-Things},'' \emph{IEEE Trans. Emerg. Topics Comput.}, vol.~5,
  no.~4, pp. 586--602, 2017.

\bibitem{pan2021iout}
X.~Pan, Y.~Shen, and J.~Zhang, ``{IoUT} based underwater target localization in
  the presence of time synchronization attacks,'' \emph{IEEE Trans. Wirel.
  Commun.}, vol.~20, no.~6, pp. 3958--3973, 2021.

\bibitem{marano2008distributed}
S.~Marano, V.~Matta, and L.~Tong, ``Distributed detection in the presence of
  {B}yzantine attacks,'' \emph{IEEE Trans. Signal Process.}, vol.~57, no.~1,
  pp. 16--29, 2009.

\bibitem{zhang2020asymptotically}
J.~Zhang and X.~Wang, ``Asymptotically optimal stochastic encryption for
  quantized sequential detection in the presence of eavesdroppers,'' \emph{IEEE
  Trans. Inf. Theory}, vol.~66, no.~3, pp. 1530--1548, 2020.

\bibitem{vempaty2013distributed}
A.~Vempaty, L.~Tong, and P.~K. Varshney, ``Distributed inference with
  {B}yzantine data: State-of-the-art review on data falsification attacks,''
  \emph{IEEE Signal Process. Mag.}, vol.~30, no.~5, pp. 65--75, 2013.

\bibitem{zhang2015Asymptotically}
J.~Zhang, R.~S. Blum, X.~Lu, and D.~Conus, ``Asymptotically optimum distributed
  estimation in the presence of attacks,'' \emph{IEEE Trans. Signal Process.},
  vol.~63, no.~5, pp. 1086--1101, 2015.

\bibitem{mothukuri2021federated}
V.~Mothukuri, P.~Khare, R.~M. Parizi, S.~Pouriyeh, A.~Dehghantanha, and
  G.~Srivastava, ``Federated-learning-based anomaly detection for {IoT}
  security attacks,'' \emph{IEEE Internet Things J.}, vol.~9, no.~4, pp.
  2545--2554, 2022.

\bibitem{zhang2018attack}
J.~Zhang, X.~Wang, R.~S. Blum, and L.~M. Kaplan, ``Attack detection in sensor
  network target localization systems with quantized data,'' \emph{IEEE Trans.
  Signal Process.}, vol.~66, no.~8, pp. 2070--2085, 2018.

\bibitem{puthal2018blockchain}
D.~Puthal, N.~Malik, S.~P. Mohanty, E.~Kougianos, and C.~Yang, ``The blockchain
  as a decentralized security framework [future directions],'' \emph{IEEE
  Consum. Electron. Mag.}, vol.~7, no.~2, pp. 18--21, 2018.

\bibitem{dinh2018untangling}
T.~T.~A. Dinh, R.~Liu, M.~Zhang, G.~Chen, B.~C. Ooi, and J.~Wang, ``Untangling
  blockchain: A data processing view of blockchain systems,'' \emph{IEEE Trans.
  Knowl. Data Eng.}, vol.~30, no.~7, pp. 1366--1385, 2018.

\bibitem{zhuang2020blockchain}
P.~Zhuang, T.~Zamir, and H.~Liang, ``Blockchain for cybersecurity in smart
  grid: A comprehensive survey,'' \emph{IEEE Trans. Ind. Informat.}, vol.~17,
  no.~1, pp. 3--19, 2021.

\bibitem{ferrag2019deepcoin}
M.~A. Ferrag and L.~Maglaras, ``Deepcoin: A novel deep learning and
  blockchain-based energy exchange framework for smart grids,'' \emph{IEEE
  Trans. Eng. Manage.}, vol.~67, no.~4, pp. 1285--1297, 2020.

\bibitem{liu2021blockchain}
H.~Liu, S.~Zhang, P.~Zhang, X.~Zhou, X.~Shao, G.~Pu, and Y.~Zhang, ``Blockchain
  and federated learning for collaborative intrusion detection in vehicular
  edge computing,'' \emph{IEEE Trans. Veh. Technol.}, vol.~70, no.~6, pp.
  6073--6084, 2021.

\bibitem{hassija2020traffic}
V.~Hassija, V.~Gupta, S.~Garg, and V.~Chamola, ``Traffic jam probability
  estimation based on blockchain and deep neural networks,'' \emph{IEEE Trans.
  Intell. Transp. Syst.}, vol.~22, no.~7, pp. 3919--3928, 2021.

\bibitem{jiang2020blockchain}
X.~Jiang, F.~R. Yu, T.~Song, Z.~Ma, Y.~Song, and D.~Zhu, ``Blockchain-enabled
  cross-domain object detection for autonomous driving: A model sharing
  approach,'' \emph{IEEE Internet Things J.}, vol.~7, no.~5, pp. 3681--3692,
  2020.

\bibitem{yang2021privacy}
Q.~Yang and H.~Wang, ``Privacy-preserving transactive energy management for
  {IoT}-aided smart homes via blockchain,'' \emph{IEEE Internet Things J.},
  vol.~8, no.~14, pp. 11\,463--11\,475, 2021.

\bibitem{zhang2017functional}
J.~Zhang, R.~S. Blum, L.~M. Kaplan, and X.~Lu, ``Functional forms of optimum
  spoofing attacks for vector parameter estimation in quantized sensor
  networks,'' \emph{IEEE Trans. Signal Process.}, vol.~65, no.~3, pp. 705--720,
  2017.

\bibitem{zhang2018approaches}
J.~Zhang, R.~S. Blum, and H.~V. Poor, ``Approaches to secure inference in the
  {Internet of Things}: Performance bounds, algorithms, and effective attacks
  on {IoT} sensor networks,'' \emph{IEEE Signal Process. Mag.}, vol.~35, no.~5,
  pp. 50--63, Sept 2018.

\bibitem{nakamoto2008bitcoin}
S.~Nakamoto, ``Bitcoin: A peer-to-peer electronic cash system,''
  \emph{Decentralized Business Review}, p. 21260, 2008.

\bibitem{zaghloul2020bitcoin}
E.~Zaghloul, T.~Li, M.~W. Mutka, and J.~Ren, ``Bitcoin and blockchain: Security
  and privacy,'' \emph{IEEE Internet Things J.}, vol.~7, no.~10, pp.
  10\,288--10\,313, 2020.

\bibitem{karame2012two}
G.~Karame, E.~Androulaki, and S.~Capkun, ``Two bitcoins at the price of one?
  double-spending attacks on fast payments in bitcoin,'' \emph{IACR Cryptol.
  ePrint Arch.}, vol. 2012, no. 248, 2012.

\bibitem{hu2019collaborative}
B.~Hu, C.~Zhou, Y.-C. Tian, Y.~Qin, and X.~Junping, ``A collaborative intrusion
  detection approach using blockchain for multimicrogrid systems,'' \emph{IEEE
  Trans. Syst., Man, Cybern. Syst.}, vol.~49, no.~8, pp. 1720--1730, 2019.

\bibitem{hassan2022anomaly}
M.~U. Hassan, M.~H. Rehmani, and J.~Chen, ``Anomaly detection in blockchain
  networks: A comprehensive survey,'' \emph{IEEE Commun. Surveys Tuts.}, pp.
  1--1, 2022.

\bibitem{jiang2021edge}
X.~Jiang, F.~R. Yu, T.~Song, and V.~C. Leung, ``Edge intelligence for object
  detection in blockchain-based internet of vehicles: Convergence of symbolic
  and connectionist {AI},'' \emph{IEEE Wireless Commun.}, vol.~28, no.~4, pp.
  49--55, 2021.

\bibitem{yang2021priscore}
Y.~Yang, Z.~Guan, Z.~Wan, J.~Weng, H.~H. Pang, and R.~H. Deng, ``Priscore:
  Blockchain-based self-tallying election system supporting score voting,''
  \emph{IEEE Trans. Inf. Forensics Secur.}, vol.~16, pp. 4705--4720, 2021.

\bibitem{hjalmarsson2018blockchain}
F.~{\TH}. Hj{\'a}lmarsson, G.~K. Hrei{\dh}arsson, M.~Hamdaqa, and
  G.~Hj{\'a}lmt{\`y}sson, ``Blockchain-based e-voting system,'' in \emph{2018
  IEEE 11th Int. Conf. Cloud Comput. (CLOUD)}, 2018, pp. 983--986.

\bibitem{lei2017blockchain}
A.~Lei, H.~Cruickshank, Y.~Cao, P.~Asuquo, C.~P.~A. Ogah, and Z.~Sun,
  ``Blockchain-based dynamic key management for heterogeneous intelligent
  transportation systems,'' \emph{IEEE Internet Things J.}, vol.~4, no.~6, pp.
  1832--1843, 2017.

\bibitem{cha2018blockchain}
S.-C. Cha, J.-F. Chen, C.~Su, and K.-H. Yeh, ``A blockchain connected gateway
  for {BLE}-based devices in the {Internet of Things},'' \emph{IEEE Access},
  vol.~6, pp. 24\,639--24\,649, 2018.

\bibitem{lombardi2018blockchain}
F.~Lombardi, L.~Aniello, S.~De~Angelis, A.~Margheri, and V.~Sassone, ``A
  blockchain-based infrastructure for reliable and cost-effective {IoT}-aided
  smart grids,'' in \emph{Living Internet Things, Cybersecur. IoT}, 2018, pp.
  1--6.

\bibitem{ling2019blockchain}
X.~Ling, J.~Wang, T.~Bouchoucha, B.~C. Levy, and Z.~Ding, ``Blockchain radio
  access network {(B-RAN)}: Towards decentralized secure radio access
  paradigm,'' \emph{IEEE Access}, vol.~7, pp. 9714--9723, 2019.

\bibitem{li2018creditcoin}
L.~Li, J.~Liu, L.~Cheng, S.~Qiu, W.~Wang, X.~Zhang, and Z.~Zhang, ``Creditcoin:
  A privacy-preserving blockchain-based incentive announcement network for
  communications of smart vehicles,'' \emph{IEEE Trans. Intell. Transp. Syst.},
  vol.~19, no.~7, pp. 2204--2220, 2018.

\bibitem{zhou2018beekeeper}
L.~Zhou, L.~Wang, Y.~Sun, and P.~Lv, ``{BeeKeeper}: A blockchain-based {IoT}
  system with secure storage and homomorphic computation,'' \emph{IEEE Access},
  vol.~6, pp. 43\,472--43\,488, 2018.

\bibitem{poor2013introduction}
H.~V. Poor, \emph{An Introduction to Signal Detection and Estimation (2nd
  ed.)}.\hskip 1em plus 0.5em minus 0.4em\relax Berlin, Heidelberg:
  Springer-Verlag, 1994.

\bibitem{shan2021poligraph}
G.~Shan, B.~Zhao, J.~R. Clavin, H.~Zhang, and S.~Duan, ``Poligraph:
  Intrusion-tolerant and distributed fake news detection system,'' \emph{IEEE
  Trans. Inf. Forensics Secur.}, vol.~17, pp. 28--41, 2022.

\bibitem{rawat2011collaborative}
A.~S. Rawat, P.~Anand, H.~Chen, and P.~K. Varshney, ``Collaborative spectrum
  sensing in the presence of {B}yzantine attacks in cognitive radio networks,''
  \emph{IEEE Trans. Signal Process.}, vol.~59, no.~2, pp. 774--786, 2011.

\bibitem{kailkhura2015distributeddetect}
B.~Kailkhura, S.~Brahma, B.~Dulek, Y.~S. Han, and P.~K. Varshney, ``Distributed
  detection in tree networks: Byzantines and mitigation techniques,''
  \emph{IEEE Trans. Inf. Forensics Secur.}, vol.~10, no.~7, pp. 1499--1512,
  2015.

\bibitem{antonopoulos2014mastering}
A.~M. Antonopoulos, \emph{Mastering Bitcoin: unlocking digital
  cryptocurrencies}.\hskip 1em plus 0.5em minus 0.4em\relax " O'Reilly Media,
  Inc.", 2014.

\bibitem{pilkington2016blockchain}
M.~Pilkington, ``Blockchain technology: principles and applications,'' in
  \emph{{Research Handbook on Digital Transformations}}.\hskip 1em plus 0.5em
  minus 0.4em\relax Edward Elgar Publishing, 2016.

\bibitem{stiawan2019investigating}
D.~Stiawan, M.~Idris, R.~F. Malik, S.~Nurmaini, N.~Alsharif, R.~Budiarto
  \emph{et~al.}, ``Investigating brute force attack patterns in {IoT}
  network,'' \emph{J. Electr. Comput. Eng.}, vol. 2019, 2019.

\bibitem{karame2015misbehavior}
G.~O. Karame, E.~Androulaki, M.~Roeschlin, A.~Gervais, and S.~{\v{C}}apkun,
  ``Misbehavior in bitcoin: A study of double-spending and accountability,''
  \emph{ACM Trans. Inf. Syst. Secur.)}, vol.~18, no.~1, pp. 1--32, 2015.

\bibitem{neshenko2019demystifying}
N.~Neshenko, E.~Bou-Harb, J.~Crichigno, G.~Kaddoum, and N.~Ghani,
  ``Demystifying {IoT} security: An exhaustive survey on {IoT} vulnerabilities
  and a first empirical look on {I}nternet-scale {IoT} exploitations,''
  \emph{IEEE Commun. Surveys Tuts.}, vol.~21, no.~3, pp. 2702--2733, 2019.

\bibitem{kaur2015energy}
N.~Kaur and S.~K. Sood, ``An energy-efficient architecture for the internet of
  things (iot),'' \emph{IEEE Syst. J.}, vol.~11, no.~2, pp. 796--805, 2017.

\bibitem{ahmed2022energy}
A.~Ahmed, S.~Abdullah, M.~Bukhsh, I.~Ahmad, and Z.~Mushtaq, ``An
  energy-efficient data aggregation mechanism for iot secured by blockchain,''
  \emph{IEEE Access}, vol.~10, pp. 11\,404--11\,419, 2022.

\bibitem{jiang2023vulnerability}
Y.~Jiang and J.~Zhang, ``Vulnerability of finitely-long blockchains in securing
  data,'' \emph{arXiv preprint arXiv:2304.09965}, 2023.

\bibitem{Cover:1991}
T.~M. Cover and J.~A. Thomas, \emph{Elements of information theory}.\hskip 1em
  plus 0.5em minus 0.4em\relax New York, NY, USA: Wiley-Interscience, 1991.

\bibitem{arora2004introduction}
J.~Arora, \emph{Introduction to optimum design}.\hskip 1em plus 0.5em minus
  0.4em\relax Elsevier, 2004.

\bibitem{luo1992convergence}
Z.-Q. Luo and P.~Tseng, ``On the convergence of the coordinate descent method
  for convex differentiable minimization,'' \emph{J. Optim. Theory Appl.},
  vol.~72, no.~1, pp. 7--35, 1992.

\bibitem{daubechies2008accelerated}
I.~Daubechies, M.~Fornasier, and I.~Loris, ``Accelerated projected gradient
  method for linear inverse problems with sparsity constraints,'' \emph{J.
  Fourier Anal. Appl.}, vol.~14, no.~5, pp. 764--792, 2008.

\bibitem{boyd2004convex}
S.~P. Boyd and L.~Vandenberghe, \emph{Convex optimization}.\hskip 1em plus
  0.5em minus 0.4em\relax Cambridge university press, 2004.

\end{thebibliography}


\end{document}